\documentclass[12pt]{article}
\pdfoutput=1
\usepackage{amsmath,amssymb,amsthm}
\usepackage{bm}
\usepackage[usenames,dvipsnames]{xcolor}
\usepackage{graphicx}
\usepackage{float}
\usepackage{array}
\usepackage{arydshln}
\usepackage{multirow}
\usepackage{rotfloat}
\usepackage{caption}
\usepackage{subcaption}
\usepackage[normalem]{ulem}
\usepackage{cite}
\usepackage{hyperref}
\usepackage{cleveref}

\graphicspath{{figures/}}

\newcommand{\ov}{\overline}

\theoremstyle{definition}

\begin{document}

\baselineskip=18pt  
\numberwithin{equation}{section}  
\allowdisplaybreaks  


\vspace*{-2cm} 
\begin{flushright}
{\tt IFT-UAM/CSIC-19-112}\\
\end{flushright}

\vspace*{0.8cm} 
\begin{center}
  {\LARGE Superconformal Blocks for Mixed 1/2-BPS Correlators with $SU(2)$
R-symmetry}

 \vspace*{1.5cm}
{Florent Baume$\,^1$, Michael Fuchs$\,^1$, Craig Lawrie$\,^2$}\\

 \vspace*{1.0cm} 
{\it $^1$
  Instituto de F\'isica Te\'orica UAM-CSIC, Cantoblanco, 28049 Madrid, Spain} \\
  {\texttt{florent.baume@uam.es, michael.fuchs@uam.es}} \\

\bigskip
{\it $^2$Department of Physics and Astronomy, University of Pennsylvania,\\
Philadelphia, PA 19104, USA} \\
  {\texttt{gmail:~craig.lawrie1729}} \\
  
\end{center}
\vspace*{.5cm}
%
\noindent 
For SCFTs with an $SU(2)$ R-symmetry, we determine the superconformal blocks
that contribute to the four-point correlation function of a priori distinct
half-BPS superconformal primaries as an expansion in terms of the relevant bosonic conformal
blocks. This is achieved by using the superconformal Casimir equation and the
superconformal Ward identity to fix the coefficients of the bosonic blocks
uniquely in a dimension-independent way. In addition we find that many of the resulting coefficients are
related through a web of linear transformations of the conformal data.

\newpage
\tableofcontents

\section{Introduction}

The use of the Operator Product Expansion (OPE) has shown to be a very powerful
tool to study Conformal Field Theories (CFTs). Being convergent
\cite{Luscher:1975js,Mack:1976pa}, it allows for a reduction of any $n$-point
correlation function to a function depending solely on the kinematic and
three-point function data.  In particular, the four-point function of scalar
fields is expanded in a series of so-called bosonic conformal blocks, depending
only on conformal invariants.  These blocks satisfy a second order differential
equation, the Casimir equation, from which an explicit form of the blocks can be derived in
even spacetime dimensions \cite{Dolan:2000ut,Dolan:2003hv}.

With the advent of the numerical bootstrap \cite{Rattazzi:2008pe}, see
\cite{Rychkov:2016iqz,Simmons-Duffin:2016gjk,Poland:2018epd,Chester:2019wfx}
for reviews, there has been a renewed interest in the study of conformal blocks.
Recent progress built on earlier works in two dimensions
\cite{Ferrara:1971vh,Migdal:1972tk,Ferrara:1973vz,Ferrara:1973yt,Polyakov:1974gs,Ferrara:1974ny,Ferrara:1974pt,Ferrara:1974nf,Dobrev:1977qv}
and the associativity property of the OPE to find bounds on the conformal
data of given unitary theories. The case where the CFT is endowed with extended
supersymmetry is furthermore of particular interest: the possible superconformal multiplets follow a
strict classification \cite{Buican:2016hpb,Cordova:2016emh} and there exists a
subset of these multiplets whose conformal dimensions are fixed by unitarity.
For such multiplets, the shortening conditions greatly simplify the structure
of the conformal blocks. With eight Poincar\'e supercharges and an R-symmetry
group containing $SU(2)_R$, results are known for conformal blocks of
four-point functions involving half-BPS operators in $3D,\,\mathcal{N}=4$
\cite{Chester:2014mea,Liendo:2016ymz}, $4D,\,\mathcal{N}=2$
\cite{Beem:2014zpa,Lemos:2015awa}, $5D,\,\mathcal{N}=1$ \cite{Chang:2017cdx} and
$6D,\,\mathcal{N}=(1,0)$ \cite{Chang:2017xmr,Bobev:2017jhk}. 

One of the most commonly studied operators for theories with eight supercharges
is the momentum map, the short multiplet containing flavour
currents\footnote{Note that in four dimensions with $\mathcal{N}=2$, there are
	additional subtleties due to a protected subsector associated to a
two-dimensional chiral algebra.\cite{Beem:2013sza}}. This enables one to find
bounds on some of the conformal data for SCFTs with flavour
\cite{Beem:2014zpa,Chang:2017cdx,Chang:2017xmr}, where there are strong
indications that the SCFTs saturating these bounds have known string theory
constructions. Therefore the conformal bootstrap might shed some light on
the relation between SCFTs and the compactification geometries. For applications
to M-theory, see e.g. \cite{Agmon:2017xes,Agmon:2019imm}.

There are two main paths usually followed to find an explicit form of these
blocks. The first is to consider the supersymmetric Ward identity
\cite{Dolan:2001tt,Dolan:2004mu,Nirschl:2004pa}, which is the most commonly
used method. The second is to solve directly the supersymmetric version of the
Casimir differential equation \cite{Bobev:2015jxa,Bobev:2017jhk} in a fashion
analogous to the method employed to find the bosonic blocks. Both these methods
allow one to formally treat the spacetime dimension as a continuous parameter.  
In this work, we extend previous results about superconformal blocks of
four-point functions of scalar superconformal primaries falling in half-BPS, or
so-called $\mathcal{D}$-type, multiplets to more general settings.  We do so
for theories with $SU(2)$ R-symmetry; without any \emph{a priori} assumption
on the particular R-charges of the external scalars; and in a
dimension-independent way, as long as $2<d\leq6$.\footnote{We exclude $d\leq2$
	as in that case some of the generators may decouple from the
superconformal algebra, see \cite{Bobev:2017jhk}.}

Our focus on theories with R-symmetry algebras isomorphic to
$\mathfrak{su}(2)$ lies in the fact that it corresponds to that of
six-dimensional $\mathcal{N}=(1,0)$ SCFTs. These theories are quite special: first
thought not to exist, it was discovered that they were related to
six-dimensional tensionless strings \cite{Seiberg:1996vs,Ganor:1996mu}, and it
has since been observed that they serve as ``master theories'' for a host of CFTs
in lower dimensions, for example this is the ethos behind the class
$\mathcal{S}$ theories that appeared in \cite{Gaiotto:2009we}. While there is no
reason why supersymmetry should be imposed, there are no known interacting
non-supersymmetric six-dimensional SCFTs and none of them have a known
Lagrangian description. Theories in six dimensions therefore offer a very nice
playground to study non-perturbative effects and relations to string theory,
for instance their connection to the swampland program \cite{Heckman:2019bzm}. 

As six is the largest dimension allowing for a superconformal algebra
\cite{Nahm:1977tg}, $\mathcal{N}=(1,0)$ representation theory provides an
overarching language encompassing lower dimensions via dimensional
reduction\footnote{We note that for $d\leq4$, the nomenclature we are using
here might not match the one the reader is familiar with. For instance, in four
dimensions, the classification of half-BPS multiplets is refined into Higgs and
Coulomb type, commonly denoted $\mathcal{E}_r$ and $\hat{\mathcal{B}}_R$
respectively \cite{Dolan:2002zh}. We refer to \cite{Cordova:2016emh} for a
dictionary between the 6D notation and lower dimensions.}.  Note that in lower
than six dimensions the R-symmetry group might enhance
due to the transverse directions in the dimensional reduction, such as the
extra $U(1)$ factor in four dimensions. We will focus here on multiplets
that are uncharged under possible additional groups so we can treat them
homogenously across dimensions. We refer the reader to
\cite{Bobev:2015jxa,Bobev:2017jhk} for more details.  Let us review the
possible multiplets allowed in theories with $SU(2)_R$ R-symmetry
\cite{Buican:2016hpb,Cordova:2016emh}.  There can exist states which are
annihilated by a subset of the supercharges.  These null states must be absent
in unitary theories, and lead to what are referred as \emph{short multiplets},
as opposed to \emph{long multiplets}, which do not have null states. It is
standard to write long multiplets as $\mathcal{L}[\Delta,\ell,J_R]$, where
$\Delta$ is the conformal dimension of the superconformal primary, $\ell$
denotes how the superconformal primary transforms as a
traceless-symmetric\footnote{In this paper we will consider only multiplets
that have a superconformal primary in a traceless-symmetric representation of
the Poincar\'e group as these will be the only contributions to the
superconformal blocks that we consider. For a general superconformal multiplet
one should replace $\ell$ with an arbitrary representation of
$\mathfrak{so}(1,d-1)$.}  representation of $\mathfrak{so}(1,d-1)$ rotations of
the Poincar\'e algebra, and $J_R$ is the charge under $SU(2)_R$. 

The different short multiplets are denoted as the $\mathcal{A}$-,
$\mathcal{B}$-, $\mathcal{C}$-, and $\mathcal{D}$-type multiplets. Unitarity
gives lower bounds on the allowed conformal dimensions, $\Delta$, of the
superconformal primaries of long multiplets, and is moreover strong enough to
completely fix the conformal dimension of the short multiplets as a function of the
other group theoretical data and the spacetime dimension. The superconformal
multiplets can be summarised as follows :
\begin{align}\label{UnitarityBounds}
	\mathcal{L}[\Delta,\ell,J_R]\,:&\quad \Delta > 2\varepsilon\; J_R + \ell + \mu\,,\nonumber\\
	\mathcal{A}[\ell,J_R]\,:&\quad \Delta = 2\varepsilon\; J_R + \ell + 4\varepsilon-2\,,\nonumber\\
	\mathcal{B}[\ell,J_R]\,:&\quad \Delta = 2\varepsilon\; J_R + \ell + 2\varepsilon\,,\\
	\mathcal{C}[J_R]\,:&\quad \Delta = 2\varepsilon\;J_R + 2\,,\nonumber\\
	\mathcal{D}[J_R]\,:&\quad \Delta = 2\varepsilon\;J_R\,,\nonumber
\end{align}
with $\varepsilon=(d-2)/2$, $\mu=2\varepsilon$ for $2<d\leq4$ and $\mu=4\varepsilon-2$ for $4\leq
d\leq6$. We stress again that this corresponds to the standard notation for $d=6$.
Indeed, $\mathcal{A}$-type multiplets correspond to the unitarity bound of long
multiplets, which for $d\leq4$ coincides with the bound of type $\mathcal{B}$.
Type $\mathcal{C}$ is unique to six dimensions and can be traced back to the presence of 
self-dual two-forms. Moreover, type $\mathcal{C}$ and $\mathcal{D}$ will appear only with
$\ell=0$, since in this work we are restricting ourselves, without loss of
generality, to only multiplets in traceless-symmetric representations of the
Poincar\'e algebra.

Some of the short multiplets may already be familiar to the reader:
$\mathcal{D}[1/2]$ and $\mathcal{C}[0]$ correspond to free hyper- and tensor
multiplets respectively, $\mathcal{B}[0,0]$ contains the energy momentum
tensor, while $\mathcal{D}[1]$ is the momentum map discussed above, containing
the conserved currents associated to possible flavour symmetries
\cite{Buican:2016hpb,Cordova:2016emh}.

This article is structured as follows: in section
\ref{sec:Structure4PtFunction} we shortly review the decomposition of
four-point functions as series of superconformal blocks written in terms of
bosonic blocks, differentiating between two possible approaches. One uses a
decomposition involving projectors onto irreducible representations of
$SU(2)_R$, while the other introduces auxiliary variables for the
R-symmetry. In section \ref{4ptConstraints} we discuss constraints the
blocks must satisfy, and how one can extract selection rules for the allowed
multiplets. More precisely, we show how the Casimir equation encodes two
different types of constraint; and how the Ward identity has to be modified to
take into account different external fields. We also comment about the
crossing symmetry these correlators must satisfy. Section \ref{sec:Results}
solves these constraints and discusses some properties of the blocks and their
coefficients. We give our conclusions in section \ref{sec:Conclusions}. In the
appendices we discuss our conventions for the superconformal group, how to
derive the $SU(2)_R$ harmonics and the Casimir differential operators. We also
review various relations satisfied by the Jack polynomials, and give a
non-exhaustive list of the coefficients of the superconformal blocks.  In
addition, we attach a \texttt{Mathematica} file to the arXiv submission of
this article containing an exhaustive list of the coefficients for all the
superconformal blocks that appear in the four-point functions of
$\frac{1}{2}$-BPS scalar operators.

\section{Structure of Four-point Functions In (S)CFTs}\label{sec:Structure4PtFunction}

In this section, we review the structure imposed by conformal invariance on
four-point functions of (super)conformal primaries. We start by recalling the
non-supersymmetric results to set our notation and conventions, and then move
to the case of superconformal primaries of $\mathcal{D}$-type multiplets. In that case we
will present two different---but equivalent---approaches, namely a
decomposition in terms of projectors of the R-symmetry, and one involving an
auxiliary variable.

In the non-supersymmetric case, the four-point functions of four \emph{a priori}
different conformal scalar primaries, $\phi_i$, of conformal dimension,
$\Delta_i$, is well known to admit a decomposition in term of bosonic conformal
blocks , $g^{\Delta_{12},\Delta_{34}}_{\Delta,\ell}$,
\cite{Dolan:2000ut,Dolan:2003hv,Dolan:2011dv}
\begin{equation}
	\left<\phi_1(x_1)\phi_2(x_2)\phi_3(x_3)\phi_4(x_4)\right> 
	= K_4 \sum_{\mathcal{O}} \lambda_{12\mathcal{O}}\lambda_{34\mathcal{O}}\,g^{\Delta_{12},\Delta_{34}}_{\Delta,\ell}(u,v)\,.
\end{equation}
The sum is taken over all conformal primaries, $\mathcal{O}$, with conformal
data, $(\Delta,\ell)$, allowed in the OPEs, and $\lambda_{ij\mathcal{O}}$
corresponds to the coefficient of the three-point function,
$\left<\phi_i\phi_j\mathcal{O}\right>$. Moreover, the kinematic prefactor will
depend on the conformal dimensions of the external primaries, $\Delta_i$, and can be shown to take
the general form 
\begin{equation} \label{conformalfactor}
	K_4 = \frac{1}{(x_{12}^2)^\frac{\Delta_1 + \Delta_2}{2}(x_{34}^2)^\frac{\Delta_3 + \Delta_4}{2} }
	\left( \frac{x_{24}^2}{x_{14}^2} \right)^\frac{\Delta_{12}}{2}
	\left( \frac{x_{14}^2}{x_{13}^2} \right)^\frac{\Delta_{34}}{2}\,,
\end{equation}
where
\begin{equation}
  x_{ij}=\left|x_i-x_j\right|  \,, \quad \text{ and } \quad \Delta_{ij} =
  \Delta_i - \Delta_j \,.
\end{equation}
The blocks are invariant under conformal transformation and therefore depend on
the two independent invariant cross-ratios, defined by
\begin{equation}\label{crossRatios}
	u = \frac{x_{12}^2 x_{34}^2}{x_{13}^2 x_{24}^2}=z\bar{z}\,,
	\qquad
	v = \frac{x_{14}^2 x_{23}^2}{x_{13}^2 x_{24}^2}= (1-z)(1-\bar{z}) \,.
\end{equation}
We have directly defined the two common variables, $(z,\bar{z})$, that will be
convenient throughout this work. We note that while in Euclidean space
these are complex conjugate, they are independent real variables for Lorentzian
signature.

The conformal blocks satisfy various properties related to crossing symmetries
of the four-point function, and can be computed as the solution of a partial
differential equation, dubbed the Casimir equation. We delay a discussion of
these properties to section \ref{4ptConstraints},
where we will delve into more details.

Here, we are interested in superconformal theories, and imposing supersymmetry
on top of conformal invariance will act as selection rules for the OPE in two
ways: first, the R-symmetry plays the role of a flavour symmetry, restricting
the possible representations allowed in the OPE of the external primaries;
second, the bosonic blocks will rearrange themselves into superconformal blocks
whose structure is compatible with superconformal representation theory. 

The remainder of this section is dedicated to outline the
selection rules and structure of four-point functions of SCFTs with eight supercharges and a R-symmetry
algebra isomorphic to $\mathfrak{su}(2)_R$.

\subsection{Four-point Functions of $\mathcal{D}$-type Superconformal primaries}\label{sec:fourPointD}

Let us now focus on the four-point function of superconformal primaries of a
$\mathcal{D}$-type multiplet belonging to an SCFT with $SU(2)$ R-symmetry. As reviewed
in the introduction, these multiplets, denoted $\mathcal{D}[J_R]$, are
half-BPS and fall into the spin-$J_R$ representation of the R-symmetry
group.\footnote{Notice that e.g. \cite{Buican:2016hpb,Cordova:2016emh} use
Dynkin indices to label the R-charge, which are integer valued. We
choose to use the spin notation---half-integer labels---in order to
unclutter many expressions.} These multiplets obey a shortening condition that
relates the conformal dimension of their superconformal primary to the R-charge
\cite{Minwalla:1997ka,Buican:2016hpb,Cordova:2016emh}:
\begin{align}\label{dimensionD}
	\Delta = 2 \varepsilon\; J_R\,,\qquad \varepsilon = \frac{d-2}{2}\,.
\end{align}
The spacetime dimension, $d$, is left arbitrary and as we will see most of the
expressions we will deal with are valid for any $d$.

The spin-$J$ representation of $SU(2)$, with $J\in \frac{1}{2}\mathbb{N}$, is
an irreducible representation that can be constructed as the $2J$th symmetric
tensor power of the fundamental representation, $\bm{2}$, that is
\begin{equation}
  Sym^{2J}\bm{2} = \bm{2J + 1} \,,
\end{equation}
where, as usual, boldface denotes an irreducible representation by its
dimension. We can realise an operator that transforms in this way by
introducing $2J$ symmetric fundamental indices
\begin{equation}
	\mathcal{O}^{(\alpha_1\dots \alpha_{2J})}(x)\,, \quad \,\alpha_i = 1,2 \,,
\end{equation}
which are raised and lowered by the usual Levi--Civita tensor,
$\varepsilon_{\alpha\beta}$, and ${(\alpha_1\dots\alpha_n)}$ indicates the
symmetrisation of the indices.
Alternatively, one can introduce an index, $M$, which runs over the spins
of the $\bm{2J+1}$ representation in which the operator is transforming. The
spins in the representation are $M = J, J - 1, \cdots, -J$. We will prefer here the
latter notation, $\mathcal{O}^M(x)$, when we consider a scalar transforming in the
spin-$J$ representation of the $SU(2)$ R-symmetry. 

Let $\phi_i^{M_i}$ be the superconformal primary of any half-BPS
superconformal multiplet, $\mathcal{D}[J_i]$, with conformal dimension, $\Delta_i$, set
by equation \eqref{dimensionD}. The correlation function of four of these
primaries is severely constrained by symmetry. First, as we saw in the
non-supersymmetric case, conformal symmetry fixes the spacetime dependence up
to a function of the invariant cross-ratios, $u,v$, and a kinematic term that
can be factored out:
\begin{equation}
	\left< \phi_1^{M_1}(x_1) \phi_2^{M_2}(x_2) \phi_3^{M_3}(x_3) \phi_4^{M_4}(x_4) \right> = K_4 \, F^{M_1M_2M_3M_4} (u,v) \,.
\end{equation}
Treating the R-symmetry as a flavour symmetry, the four-point function must
be an invariant tensor of $SU(2)$, and therefore $F^{M_1M_2M_3M_4}(u,v)$ is an invariant
under both conformal symmetry and $SU(2)_R$. Using the OPE, this function
can be expanded into contributions coming from each of the superconformal
multiplets, $\chi$, allowed in the expansion,
\begin{equation}
	F^{M_1M_2M_3M_4}(u,v) = \sum_\chi \lambda_{12\chi}\lambda_{34\chi} \mathcal{G}_\chi^{M_1M_2M_3M_4} (u,v)\,.
\end{equation}
Inside of each superconformal multiplet there are primary operators that
transform in different representations of the $SU(2)$ R-symmetry. By
introducing projectors on the spin-$J$ representation, $P_{J}^{M_1M_2M_3M_4}$,
in the superconformal block, $\mathcal{G}_\chi^{M_1M_2M_3M_4}$, we can further
split the four-point function into a sum over all allowed R-symmetry channels.
Thus for each superconformal multiplet we can expand as
\begin{equation}\label{exp1}
	\mathcal{G}_\chi^{M_1M_2M_3M_4} = \sum_{J \in \mathcal{J} } P_{J}^{M_1M_2M_3M_4}\; \mathcal{G}_\chi^{J} (u,v) \, ,
\end{equation}
where $\mathcal{J}$ is the set of all allowed propagating spins. By considering
the $s$-channel for the OPE one can see that the set of $SU(2)$ representations
that correspond to the propagating spins is determined by the tensor products
\begin{equation}
	 ((\bm{2J_1 + 1}) \otimes (\bm{2J_2 + 1})) \cap ((\bm{2J_3 + 1})
	\otimes (\bm{2J_4 + 1})) \,.
\end{equation}
Recalling how $SU(2)$ tensor products decompose, one can easily compute
the set of propagating spins $\mathcal{J}$ for any given $J_1, \cdots, J_4$.
This can be written as  
\begin{equation}\label{Jrange}
	\mathcal{J} = \left\{\textrm{Max}(|J_2 - J_1| , |J_4 - J_3|),\,
  \dots\,,\,\text{Min}(J_1+J_2,J_3+J_4) \right\} \,,
\end{equation}
where we add the caveat that the set is empty if the start and end values
different by $n + \frac{1}{2}$ for some integer $n$---in such a case there are no
propagating spins.
To give an explicit example, for coinciding representations, 
$J = J_i$, then we have
\begin{equation}
  \mathcal{J} = \{ 2J, 2J - 1, \cdots, 0 \} \,,
\end{equation}
as the propagating spins appearing in the sum in (\ref{exp1}).

Finally, the superconformal blocks can be decomposed into bosonic blocks,
where by symmetry only the bosonic components of a superconformal multiplet
can contribute to the OPE. In fact, since we are considering OPEs of scalar
fields, only fields that are symmetric and traceless are allowed. Therefore
the spin-$J$ part of superconformal block, $\mathcal{G}_\chi^{J}(u,v)$,
associated to the superconformal multiplet can be written as a sum over bosonic
conformal blocks, $g^{\Delta_{12}, \Delta_{34}}_{\Delta, \ell}(u, v)$, and
collects the contribution of all the constituent (non-supersymmetric) primary
fields of the superconformal multiplet that have the aforementioned spin, $J$:
\begin{equation}\label{Fexp}
	\mathcal{G}_\chi^{ J}(u,v) =  \sum_{ (\tilde{\Delta}, \tilde{\ell}) \in \chi} f^J_{\tilde{\Delta} , \tilde{\ell}}\,g_{\tilde{\Delta}, \tilde{\ell}}^{\Delta_{12}, \Delta_{34}}(u,v) \,,
\end{equation}
where $(\tilde{\Delta},\tilde{\ell})$ correspond to the data of the relevant
superconformal descendants inside the multiplet with fixed R-charge, $J$.
Finding the explicit expression for a superconformal block therefore reduces to
determining the coefficients $f_{\Delta,\ell}^J$. In order to do
so, we will use both the Casimir equation and Ward identity to constrain them,
and eventually fix them all in terms of the data of the superconformal primary.

The full superconformal block associated to a multiplet, $\chi$, whose primary
has conformal data $(\Delta,\ell,J_R)$ can therefore be decomposed into a sum
over all $\mathcal{G}_\chi^J$. We note that throughout this paper the
labelling $(\Delta,\ell,J_R)$ will always denote the data of the superconformal primary of a
given superconformal multiplet, $\chi$, while $(\tilde{\Delta},\tilde{\ell},J)$
will refer to that of any of its states, including the primary. In the case of
eight supercharges, one can show that the allowed R-charges inside a
multiplet are between $J_R-2$ and $J_R+2$. Furthermore each application of a
supercharge will raise the conformal dimension by $\frac{1}{2}$, and possibly
change its Poincar\'e representation, thus the superconformal block can be
written as
\begin{equation}\label{BlockUncontracted}
	\mathcal{G}^{M_1M_2M_3M_4}_\chi = \sum_{J=J_R-2}^{J_R+2}\sum_{m = 0}^4\sum_{n=-2}^2P_J^{M_1M_2M_3M_4} f^J_{\Delta+m,\ell+n} g_{\Delta+m,\ell+n}^{\Delta_{12},\Delta_{34}}(u,v)\,.
\end{equation}
Of course, depending on the type of multiplet considered and its content, not
all $f^J_{\Delta,\ell}$ are non-vanishing, and some of them can be set to zero by
group theoretical arguments. We will further expand on the structure of
superconformal multiplets in section \ref{4ptConstraints}.

\subsection{R-symmetry Variables}\label{sec:RsymmetryVariables}

It is sometimes useful to introduce auxiliary variables, $Y^\alpha$, to encode
R-symmetry transformations in a more convenient way
\cite{Nirschl:2004pa,Dolan:2004mu}. These variables can be used to contract all
possible R-symmetry indices of a given operator, 
\begin{equation}
	\mathcal{O}(x, Y) = Y_{\alpha_1} \cdots Y_{\alpha_{2J}} {\cal O}^{\alpha_1 \dots \alpha_{2J}} (x)\,,
\end{equation}
such that if $\mathcal{O}(x)$ is a scalar primary of conformal dimension,
$\Delta$, in the spin-$J$ representation, $\mathcal{O}(x,Y)$ is a homogeneous
function of degree $(-\Delta,2J)$. 

In the case of four-point functions of $\mathcal{D}$-type primaries, the
discussion at the beginning of this section has to be modified to take into
account auxiliary variables, and an additional prefactor related to
$Y^\alpha$ can be extracted from the correlator, 
\begin{equation}\label{fourpt}
	\left< \phi_1(x_1, Y_1)\phi_2(x_2, Y_2)\phi_3(x_3, Y_3)\phi_4(x_4, Y_4)  \right> =  K_4 \,K_4^{R} F(u,v;w) \, .
\end{equation}
The quantity $K_4$ is the usual kinematic prefactor given by equation
\eqref{conformalfactor}, while $K_4^R$ takes into account the homogeneity of
the four-point function with respect to $Y^\alpha$.
\begin{equation}
	K_4^R=
	\left(Y_{12}\right){}^{a_1} \left(Y_{13}\right){}^{a_2}
	\left(Y_{14}\right){}^{-a_1-a_2+2 J_1}
	\left(Y_{23}\right){}^{-a_1-a_2+J^+_{12}+J_{34}} 
	\left(Y_{24}\right){}^{a_2-J_{12}-J_{34}} 
	\left(Y_{34}\right){}^{a_1-J^+_{12}+J_{34}^+}\,.
\end{equation}
where we defined the quantity $J_{ij} = J_i - J_j$ and $J^+_{ij} = J_i + J_j$.
The so-far undefined function in the RHS of \eqref{fourpt} must be an invariant
under both conformal and R-symmetry transformations, and we must therefore find
the analogue of the invariant cross-ratios $u,v$ for the R-symmetry. It can be shown that
the unique candidate is given by 
\begin{equation}
	w = {(Y_1 \cdot Y_2) (Y_3 \cdot Y_4)  \over (Y_1 \cdot Y_4) (Y_2 \cdot Y_3) }\,,\qquad Y_{ij} =  Y_i^{\alpha} Y_j^\beta \varepsilon_{\alpha \beta}\,.
\end{equation}
The coefficients, $a_1,\, a_2$, in the prefactor are arbitrary constants that
effectively rescale $F(u,v;w)$ by factors of $w$ and $1+w$, respectively. A
particular choice for these constants is merely a choice of convention; for more details on the possible
choices and how they relate to previous
works see appendix \ref{app:ConformalFrame}.

In a similar fashion that was reviewed in the case of ``uncontracted'' fields,
the invariant function can be split into contributions from each
superconformal multiplet, $\chi$, 
\begin{equation}
	F(u,v;w) = \sum_\chi \lambda_{12\chi}\lambda_{34\chi} \;\mathcal{G}_\chi(u,v;w) \,,
\end{equation}
and into contributions from the different R-symmetry channels given now by
$SU(2)_R$ harmonics,
\begin{equation}\label{superconformalBlockHarmonics}
	\mathcal{G}_\chi(u,v;w) = \sum_{J \in {\cal J}}  \,\, \mathcal{P}_{J}^{J_{12},J_{34}}(w)  \,\,  \mathcal{G}_{J}^\chi(u,v) \, .
\end{equation}
The harmonics, $\mathcal{P}_J(w)$ are obtained by inserting the quadratic
Casimir for $SU(2)_R$ in the four-point function, in a similar way to what is
usually done to obtain bosonic conformal blocks, 
\begin{equation}\label{hyperGeometricHarmonic}
	\mathcal{P}_{J}^{J_{12},J_{34}}(w) =c_J \frac{w^{-J-(a_1-(J_1+J_2))}}{ (1+w)^{a_2}} \, _2F_1\big(-(J+J_{12}),-(J+{J_{34})};-2 J;-w\big) \,,
\end{equation}
and are related to the R-symmetry projectors, $P^{M_1M_2M_3M_4}_J$,
introduced in \eqref{exp1}. Note that the $SU(2)_R$ harmonics \emph{a priori}
depend on the combination $J_1 + J_2$. Our choice of normalisation, $a_1 = J_1
+ J_2\,,a_2=J_{34}$, absorbs it and leaves the dependence on external data only
on the difference $J_{ij}=J_i - J_j$. The full four-point function does of
course not depend on this convention, but it will make some of the intermediate
expressions easier. We will comment as to possible differences between
conventions when needed.

The hypergeometric function can, in principle, be recast into, perhaps more
familiar, Jacobi polynomials \cite{Nirschl:2004pa}, but in practical
computations we find the hypergeometric function more convenient. We refer to
appendix \ref{app:ConformalFrame} for additional details on the derivation of
\eqref{hyperGeometricHarmonic} and possible conventions. 

The quantity $c_J$ is an arbitrary constant we choose to be 
\begin{equation}
	c_J =  \frac{(-1)^{J} \Gamma (2 J+1)}{\left(1-J_{12}\right)_{J} \left(1-J_{34}\right)_{J}} \,,
\end{equation}
such that the spin-$J$ contribution to the superconformal
blocks, $\mathcal{G}_J^\chi$, defined here agrees with those of the
``uncontracted'' notation when using our choice of normalisation for the
projectors (see later in equation \eqref{convention}). Using the expansion
\eqref{Fexp} in terms of bosonic blocks one finds the contribution to the
four-point function of a superconformal multiplet, $\chi$, is
\begin{equation}\label{superconformalblock}
	\mathcal{G}_\chi(u,v; w) = \sum_{J \in \mathcal{J}}\sum_{m,n} f^J_{
  \Delta+m, \ell+n}\;  \mathcal{P}_{J}^{J_{12}, J_{34}}(w)\;   g_{\Delta+m, \ell+n}^{\Delta_{12}, \Delta_{34}}(u,v) \, .
\end{equation}

As the set of different possible superconformal multiplets is known, we
can further use the structure of these multiplets as a selection rules for the
possible multiplets appearing in the OPE. As we will see in the following
sections, the Casimir and Ward identities can in general only be satisfied if
\emph{all} 
the non-zero bosonic blocks in the decomposition \eqref{Fexp} are
present---in the case of coincident $J_i$ some of them are vanishing, but their
absence can be traced back to crossing symmetry. 

This constrains further the allowed superconformal multiplets in the expansion. Indeed,
let us denote the largest spin in \eqref{Jrange} by $J_\text{max}$. A long
multiplet whose superconformal primary has R-charge $J_R=J_\text{max}$ also
contains states with $J=J_\text{max}+1\,,J_\text{max}+2$. $SU(2)_R$ symmetry
prohibits these states to appear in the OPE and, as we will see, the full
multiplet is either forbidden to participate, or the structure of the
coefficients $f^{J>J_\text{max}}_{\tilde{\Delta},\tilde{\ell}}$ is such that
they precisely vanish. In fact, the information about possible null states,
e.g. the conservation of flavour currents inside short multiplets
$\mathcal{D}[1]$, is also encoded in the structure of the coefficients.
The case of $\mathcal{D}$-type multiplets is not plagued by this constraint, as
the descendants have an R-charge smaller than that of the superconformal primary, but for the
$\mathcal{B}$-type multiplet the presence of a state with R-charge $J_R+1$
leads to reduced options.

The last two possible types of short multiplets, $\mathcal{A}$
and $\mathcal{C}$ can be shown to be incompatible with the Ward identity and do
not contribute to the four-point function \cite{Ferrara:2001uj,Chang:2017xmr}. Schematically, the block
decomposition thus takes the form
\begin{equation}\label{blockDecomposition}
	F(u,v;w) \sim \sum_{J=J_\text{min}}^{J_\text{max}-2}\mathcal{L}[\Delta,\ell,J]
	+ \sum_{J=J_\text{min}}^{J_\text{max}-1}\mathcal{B}[\ell,J]
	+ \sum_{J=J_\text{min}}^{J_\text{max}}\mathcal{D}[J]
	\,,
\end{equation}
where a sum over all possible $\Delta,\ell$ allowed by unitarity
\eqref{UnitarityBounds} 
is understood.	When the R-symmetry group contains an extra factor, such as
$\mathcal{N}= 4$ in three dimensions where it is semi-simple,
$SU(2)\times SU(2)$, this selection rule is modified. In particular,
for $\mathcal{N}=2$ in four dimensions, the presence of the abelian
group, $U(1)$, implies the existence of an extra sector with
contributions from superconformal multiplets with a primary in a
non-traceless-symmetric representation of the Lorentz group
\cite{Beem:2014zpa}.

We end this section by noting that while subsection \ref{sec:fourPointD} is up
to minor modifications similar to the decomposition used in
\cite{Bobev:2015jxa} to obtain the blocks of four momentum-map operators, the
decomposition with auxiliary variables gets more involved when considering
non-coincident operators, and depends heavily on $J_{12}$ and $J_{34}$. When
setting them to zero, the $SU(2)_R$ harmonic \eqref{hyperGeometricHarmonic}
reduces to Legendre polynomials and one recovers the familiar expressions used
in e.g. \cite{Beem:2014zpa,Chang:2017xmr}.

\section{Constraints on Four-point Functions of Half-BPS Primaries}\label{4ptConstraints}

Before delving into the constraints satisfied by blocks and, by extension, the
coefficients $f^J_{\tilde{\Delta},\tilde{\ell}}$, let us recall some facts
about the representation theory of the superconformal group with extended
supersymmetry that will prove useful when computing the superconformal blocks.

Unitary representations of these groups have been extensively studied, starting
with \cite{Dobrev:1985qv,Dobrev:1985vh,Dobrev:1985qz}, and more recently with
\cite{Minwalla:1997ka, Dolan:2001tt, Bhattacharya:2008zy, Cordova:2016emh,
Buican:2016hpb, Cordova:2016xhm}. In addition to the Poincar\'e generators,
there are additional fermionic generators, $Q_{\alpha A},\, S^{\alpha A}$---in
our case eight of each---which act in conjunction with $P_\mu$ and $K_\mu$ as
ladder operators. Our convention for the superconformal algebra is set in
appendix \ref{app:superconformalGroup}. 

A superconformal primary, $\mathcal{O}$ with conformal dimension, $\Delta$,
R-charge, $J_R$, and falling into a traceless-symmetric representation of the
Poincar\'e group\footnote{There are of course multiplets in other
  representations of the Poincar\'e group, but these will not contribute to
the quantities computed in this work.}, $\ell$, is by definition annihilated
by all conformal supercharges, $S^{\alpha A}\left|\mathcal{O}\right>=0$, as
well as by the special conformal transformation generator,
$K_\mu\left|\mathcal{O}\right>=0$.  Using combinations of all (Poincar\'e)
supercharges, $Q_{a\alpha}$, we can reach an additional set of states whose
superconformal data are related to that of the superconformal primary. More
precisely, applying a supercharge to the superconformal primary will give rise
to a state whose conformal dimension has been raised by $\frac{1}{2}$, and
whose R-charge and Poincar\'e representation are changed. Each multiplet
contains primary states with, at most, conformal dimension $\Delta+4$, R-charge
between $J_R-2$ and $J_R+2$, and, focusing on traceless-symmetric
representations, Poincar\'e representation between $\ell-2$ and $\ell+2$
\cite{Buican:2016hpb,Cordova:2016emh}.

Short multiplets are then multiplets for which some descendants are annihilated
by given combinations of the supercharges. In particular, the
$\mathcal{D}$-type multiplets that are the focus of this work have a primary
annihilated by half of the supercharges and are therefore half-BPS states.
This property reveals itself crucial when studying their four-point functions,
as it leads to simplifications that do not occur in the other cases.

This section is dedicated to the constraints satisfied by the superconformal
blocks. We will first work out the constraints from the Casimir equations. One will be a differential equation while other will give a selection rule for the multiplets at
the boundary of the set of the $SU(2)_R$ representations, $\mathcal{J}$, appearing in the
OPE. We will then consider the consequences of the Ward identity when
considering the
R-symmetry auxiliary variables. This constraint will prove the strongest and
will uniquely fix all possible superconformal blocks. Lastly we will outline constraints from crossing symmetry. 

\subsection{The Superconformal Casimir Equation}\label{sec:casimirEquation}

Having set up the decomposition of the four-point function of 
$\mathcal{D}$-type superconformal multiplets into conformal blocks, we are now
ready to find how to use the Casimir equation to constrain the coefficients of
the superconformal blocks and eventually fix some of them. This technique has
already been explored in \cite{Bobev:2015jxa,Bobev:2017jhk} for theories with
four and eight supercharges, and this section generalises their results to
arbitrary R-charge for SCFTs with eight supercharges.

Before moving to the supersymmetric case, it is useful to recall how the
Casimir equation has been used to find the bosonic blocks for arbitrary
$\varepsilon$. 

The idea is to insert the completeness relation of projectors, $P_\mathcal{O}$,
and the conformal Casimir\footnote{See appendix \eqref{app:superconformalGroup}
for a definition in terms of the generators, and appendix \ref{app:ConformalFrame} for a derivation of a Casimir equation in the case of $SU(2)_R$ harmonics.}, $C^2_\text{bos}$, into the
correlator to obtain the contribution from every primary, $\mathcal{O}$, and
its descendants,
\begin{equation}
	P_{\cal O} =\sum_{a,b=(P_\mu)^n\mathcal{O},n\geq0}\left| a \right> \left< b\right|\left<a\middle|b\right>^{-1}\,,
	\qquad \mathbf{1}=\sum_{\mathcal{O}}P_\mathcal{O}\,.
\end{equation}
Letting $C^2_\text{bos}$ act on $\phi_1\phi_2$ as a first-order linear
differential operator, one finds that the bosonic blocks satisfy the
differential equation \cite{Dolan:2003hv},
\begin{equation}\label{bosonicCasimirEquation}
	\mathcal{D}_\text{bos}\, g^{\Delta_{12},\Delta_{34}}_{\Delta,\ell}(z,\bar{z}) 
	= c^\text{bos}_{\Delta,\ell}~g^{\Delta_{12},\Delta_{34}}_{\Delta,\ell}(z,\bar{z})\,.
\end{equation}
The second-order differential operator, $\mathcal{D}_\text{bos}$, depends
explicitly on the difference between the conformal dimensions of the
primaries,
\begin{align}\label{bosonicCasimirOperator}
	\mathcal{D}_\text{bos} &= \mathcal{D}_z+\mathcal{D}_{\bar{z}} +4\varepsilon \frac{z\bar{z}}{z-\bar{z}}\bigg( (1-z)\partial_z -(1-\bar{z})\partial_{\bar{z}} \bigg)\,,\\
	\mathcal{D}_z  &= 2z^2(1-z)\partial^2_z-(2+\Delta_{34}-\Delta_{12})z^2\partial_z+\frac{\Delta_{12}\Delta_{34}}{2}z\,.
\end{align}
Notice that we are considering bosonic blocks in terms of the variables
$z\,,\bar{z}$, rather that the conformal cross-ratios $u\,,v$. This will make
the various differential operators in this section simpler.

The RHS of equation \eqref{bosonicCasimirEquation} depends solely on the
quadratic Casimir eigenvalue associated to the bosonic primary with conformal
data $(\Delta,\ell)$, 
\begin{equation}\label{bosonicCasimirEigenvalue}
	c^\text{bos}_{\Delta,\ell} = \Delta(\Delta-2(\varepsilon+1))+\ell(\ell+2\varepsilon)\,.
\end{equation}
For even dimensions, equation \eqref{bosonicCasimirEquation} simplifies and an
analytic expression can be found in terms of hypergeometric functions
\cite{Dolan:2003hv}. For arbitrary $\varepsilon$ the solutions of the Casimir
equation are unknown, but there exists rapidly converging power series in terms
of radial coordinates \cite{Hogervorst:2013sma,Costa:2016xah}, as well as
recursion relations by studying their analytic structure
\cite{Kos:2013tga,Kos:2014bka,Penedones:2015aga}.

To obtain the supersymmetric version of \eqref{bosonicCasimirEquation} for
superconformal blocks, $\mathcal{G}_{\Delta,\ell}^{M_1M_2M_3M_4}$, the
conformal Casimir, its eigenvalue and projectors are replaced by
their supersymmetric cousins \cite{Bobev:2017jhk},
\begin{align}
	C^2 =& C^2_\text{bos} + C^2_\text{SUSY} + C^2_R\,,\\
	c_{\Delta,\ell,J_R} =& c^\text{bos}_{\Delta,\ell} + 4\Delta + 2\varepsilon J_R(J_R+1)\,. \label{supersymmetricCasimir}
\end{align}
The procedure to get the Casimir equation is the same as in the bosonic case,
with the exception that the additional generators have to be taken into account:
\begin{equation}\label{superCasimirEquation}
	\left(\mathcal{D}_\text{bos} + \mathcal{D}_\text{SUSY} + \mathcal{D}_R \right)\mathcal{G}_\chi^{M_1M_2M_3M_4}(z,\bar{z}) = c_{\Delta,\ell,J_R}\, \mathcal{G}^{M_1M_2M_3M_4}_\chi(z,\bar{z})\,.
\end{equation}
The quadratic Casimir, acting as differential operator, will behave differently
depending on the values of $M_i$. Note that for arbitrary values of $M_i$,
inserting the part of the Casimir involving the supercharges,
$C^2_\text{SUSY}$, in the four-point function will produce four-point
functions involving mixed scalar and fermionic fields, and the resulting
Casimir equations will not be partial differential equations. This makes it
challenging at best to compute the Casimir equation in terms of
well-defined quantities.

In the case of short $\mathcal{D}$-type superconformal multiplets however, the
primaries in the highest and lowest weight representations of the R-symmetry
are annihilated by half of the supercharges, $Q_{\alpha A}$,
respectively\cite{Minwalla:1997ka,Buican:2016hpb,Cordova:2016emh},
\begin{equation}\label{halfBPSCondition}
	Q_{1A}\phi^J = 0\,,\qquad
	Q_{2A}\phi^{(-J)}= 0\,,\qquad
	\forall\, A\,.
\end{equation}
Therefore when the superconformal block, $\mathcal{G}_\chi^{M_1M_2M_3M_4}$,
involves highest and lowest weights of the $SU(2)_R$ representations, we obtain
a well-defined partial differential equation. Without loss of generality, we
will henceforth assume that $J_1\leq J_2\leq J_3\leq J_4$. This makes the
possible values of the set $\mathcal{J}$ more tractable and makes it easier to see
when the projectors are vanishing. As shown in appendix
\ref{app:superconformalcasimir}, using the procedure introduced in
\cite{Bobev:2015jxa}, this leads us to consider two types of correlation
functions, which we dub type (I) and (II):
\begin{equation}
  \text{(I)}:\,\,\left<\phi_1^{J_1}(x_1)\phi_2^{J_2}(x_2)\phi_3^{M_3}(x_3)\phi_4^{M_4}(x_4)\right>\,,\qquad
  \text{(II)}:\,\,\left<\phi_1^{J_1}(x_1)\phi_2^{M_2}(x_2)\phi_3^{J_3}(x_3)\phi_4^{M_4}(x_4)\right>\,.
\end{equation}
The differential operator, $\mathcal{D}_\text{SUSY}$, acts differently on the
conformal blocks depending on the type of correlation function considered,
\begin{align}
  \text{(I)}: &~\mathcal{D}_\text{SUSY}  = 4(\Delta_1+\Delta_2)\,,\\
  \text{(II)}: &~\mathcal{D}_\text{SUSY}  = 4z(1-z)\partial_z+\bar{z}(1-\bar{z})\partial_{\bar{z}} - 2(z+\bar{z}) \Delta_{34}\,.\label{typeIIDSUSY}
\end{align}

In both cases, the differential operator associated to $C^2_R$ can be obtained
simply by applying it on the projector instead of $\phi_1\phi_2$. It then gives
the eigenvalue of the R-charge\footnote{Note that in the presence of an extra
R-symmetry factor, the Casimir gets additional
contributions, changing the form of the blocks. In
particular, abelian factors may allow for multiplets whose
superconformal primary is not in a traceless-symmetric representation
of the Lorentz group, see e.g. \cite{Bobev:2015jxa}.}of the bosonic block
considered,
\begin{equation}
	\mathcal{D}_R = -2\varepsilon J(J+1)\,.
\end{equation}
The different types of correlators will lead to two different types of
constraints, which we now outline. 

\subsubsection*{Constraints from Type (I)}

For the first case, type (I), the only value of $J$ leading to a non-trivial
projectors is $J=J_1+J_2$. This can be seen by writing the projectors in terms
of Clebsch--Gordan coefficients. As there is only one contribution in equation
\eqref{superCasimirEquation}, the projector drops out of the Casimir equation
and one obtains
\begin{equation}
	\bigg(\mathcal{D}_\text{bos} -2\varepsilon(J_1+J_2)(J_1+J_2-3)\bigg)\mathcal{G}_\chi^{J_1+J_2} = c_{\Delta,\ell,J_R}\mathcal{G}_\chi^{J_1+J_2}\,.
\end{equation}
Decomposing the multiplet in terms of bosonic blocks with conformal data shifted away from
the primary, $(\tilde{\Delta},\tilde{\ell})=(\Delta+m,\ell+n)$, one finds the constraint 
\begin{equation}\label{typeIConstraint}
	(m-2)(2 \Delta+m) + n(2l + n) 
	+2\varepsilon  \left(J_R(J_R+1)-J_\text{max}(J_\text{max}-3) +n-m\right) = 0\,.
\end{equation}
This constraint must be satisfied by any states in a given multiplet
susceptible to have $J=J_{\text{max}}$, and one must check whether a state is
allowed or not on a case-by-case basis. We will do so in section \ref{sec:Results}.

\subsubsection*{Constraints from Type (II)}

For correlation functions of type (II), the part of the Casimir associated to
the supercharges no longer act as multiplication, but involves derivatives with
respect to the coordinates $z,\,\bar{z}$. Moreover the projectors are no longer
vanishing and we must consider a sum over all possible R-charges in the set
$\mathcal{J}$.

The canonical normalisation for projectors is usually fixed such
that the trace of the projector gives the dimension of the representation.
Here, we have chosen a slightly different convention that will lead to a simpler
expression for the Casimir equations. Indeed, for type (II) correlators,
$M_1\,,M_3$ are fixed to be the highest weights of their respective
representations. The remaining two can be parametrised as the deviation from
the lowest weights of $J_1$ and $J_3$, $M_2 = -J_1 + m\,, M_4 = -J_3 -m$ since, by
$SU(2)$ invariance, it is required that $\sum_i M_i = 0$. Note that for coincident external
R-charges the only non-vanishing possibility is $m=0$.

As one can check, the four-point functions for any choice of $m$ are proportional
to each other. Furthermore for all allowed intermediate R-charges, $J \in
\mathcal{J}$, the projectors are non-vanishing if $m$ is chosen in the allowed
range. We thus choose our normalisation convention in such a way that all
contributions from the projectors are one for $m=0$,\footnote{We note that,
should the reader prefer using another, more canonical, convention, our
results for the superconformal blocks can be converted into any other
convention for the projectors by rescaling all the coefficients,
$f^J_{\tilde{\Delta},\tilde{\ell}}$, according to $f^J_{ \tilde{\Delta}, \tilde{\ell}}
\rightarrow f^J_{ \tilde{\Delta}, \tilde{\ell}}/P_{J }^{J_1\,(-J_1)\, J_3\,
(-J_3)}$.}
\begin{equation}\label{convention}
	P_{J}^{J_1\,(-J_1)\, J_3\, (-J_3)}= 1\, , \qquad \forall\,J \in {\cal J}\,.
\end{equation}
In this convention, the type (II) Casimir equation reduces to 
\begin{equation}\label{superCasimirEquationTypeII}
	\sum_{J\in\mathcal{J}}\bigg(\mathcal{D}_\text{bos} + \mathcal{D}_\text{SUSY} - 2\varepsilon J(J+1)-c_{\Delta,\ell,J_R} \bigg)\mathcal{G}_\chi^J(z,\bar{z}) = 0\,.
\end{equation}

Because of the derivative in \eqref{typeIIDSUSY}, the blocks, superconformal
or otherwise, are no longer eigenfunctions of $\mathcal{D}_\text{SUSY}$, and
constraints can no longer be solved block by block. To overcome this, the
usual strategy is to decompose the bosonic blocks in terms of orthogonal Jack
polynomials, $P^{\varepsilon}_{\lambda_1,\lambda_2}$
\cite{Bobev:2015jxa,Bobev:2017jhk}, which we will accomplish in section
\ref{sec:Results}.

\subsection{The Ward Identity}\label{sec:WardIdentity}

In the previous subsection, we obtained constraints on the coefficients,
$f^J_{\tilde{\Delta},\tilde{\ell}}$ by letting the Casimir operator act as a
differential operator on the four-point function, which differs depending on
the type of projectors considered. One can ask whether the same kinds of
constraints can be obtained using the auxiliary variables, $Y^\alpha$, reviewed
in section \ref{sec:RsymmetryVariables}. In this case, the constraints are
obtained through the Ward identity satisfied by the correlators
\cite{Nirschl:2004pa, Dolan:2004mu}. For half-BPS multiplets in theories with
$N_S$ superPoincar\'e supercharges, its origin lies in the fact that in
superspace, the four-point function depends on $4\times\frac{N_S}{2}=2N_S$
independent Grassmann variables. As $2N_S$ is also the total number of
fermionic generators of superconformal algebra, $Q$ and $S$, there exists a
frame in which all Grassmann variables can be set to zero. 

Vice versa this means that one can always find a superconformal completion from
the four-point function of the superconformal primaries and restore the full
Grassmann variable dependence. This completion should be well behaved as long
as we keep the spacetime coordinates apart.  It turns out that the
superconformal transformation of the coordinates are singular, and for the
four-point function to be well defined, these singularities needs to be cured,
leading to the so-called Ward identity.

Let us shortly recall the linearised superconformal transformations, $\delta$,
with $N_S=8$ \cite{Dolan:2004mu}.\footnote{We would like to warn the reader
than in \cite{Dolan:2004mu}, the use of $z$ and $\chi$ are exchanged.
We use the present notation as it has become the standard in the recent
literature.} In a frame parameterised by a pair of conformal coordinates,
$(\chi(z),\bar{\chi}(\bar{z}))$---see appendix \ref{app:ConformalFrame} for
details---these transformations induce the following pole:
\begin{equation}
	\delta_Q w = \varepsilon\, \delta_Q \chi \sim \frac{1}{w-\chi}\,,\qquad
	\chi = \frac{z}{1-z}\,,
\end{equation}
where only the pole was kept for simplicity. We refer to the original work for
more details.

A transformation of the full four-point function will therefore become singular
as the auxiliary invariant, $w$, approaches the conformal coordinate, $\chi$.
It is then clear that physical quantities should be free of such a singularity,
which can be attained by demanding that
\begin{equation}
	\left.\bigg((\partial_\chi + \varepsilon \partial_w)\left< \phi_1(x_1, Y_1)\phi_2(x_2, Y_2)\phi_3(x_3, Y_3)\phi_4(x_4, Y_4)  \right>\bigg)\right|_{\chi=w}=0\,.
\end{equation}
A similar Ward identity can also be obtained for the other variable,
$\bar{\chi}(\bar{z})$ by exchanging $\chi\leftrightarrow\bar{\chi}$. A
convenient choice of convention for the prefactor reduces it to a particularly
appealing form:
\begin{equation}
	\left.\bigg((\partial_\chi + \varepsilon \partial_w)F(u,v;w)\bigg)\right|_{\chi=w}=0\,,\qquad a_1= J_1+J_2\,,a_2=J_{34}\,.
\end{equation} In the case of arbitrary $a_1\,,a_2$ however, one must take
particular care of the prefactors $K_4\,,K_4^R$. Upon defining the variable $\alpha=(1+w)/w$, the
superconformal Ward identity takes the form 
\begin{equation}\label{WardIdentity}
	\bigg( 
	-z \varepsilon 
	\left((a_1-J_1-J_2)+z(a_2 -J_{34})\right)
	+(z-1) 
\left(z^2 \partial_z -\varepsilon \partial_\alpha \right) \bigg) F(z, \ov z, \alpha) |_{\alpha = \frac{1}{z}} = 0\,.
\end{equation}
One can then plug in the expression of the superconformal blocks and their bosonic contributions to get explicit constraints on the coefficients $f^J_{\tilde{\Delta},\tilde{\ell}}$. After the dust settles, the resulting identity depends only on $z$. As described in appendix \ref{app:ConformalFrame} the factor containing $a_1, a_2$ actually cancels against factors contained in the harmonics. As a result all $a_1, a_2$ dependence of the above Ward identity is merely a relic from the choice of $K_4^R$ and drops out in actual computations. Furthermore the resulting Ward identity applied on a specific superconformal block will factor out all contributions of the form $(J_1+J_2)$, and all the external data will boils down to the specific combinations $J_{12}\,,J_{34}$. 

\subsection{Crossing Symmetry Constraints}\label{sec:CrossingSymmetry}

Finally, we close this section on the superconformal blocks by giving the
crossing symmetries that they must satisfy. Until here, we have tacitly assumed
that the OPE leading to the block decomposition was performed in the
$s$-channel, that is to say by performing an OPE on $\phi_1$ with $\phi_2$ and
on $\phi_3$ with $\phi_4$ respectively. The other two channels, $u$ and $s$,
should be the same by the associativity properties of the OPE, and have been
recently used in the numerical bootstrap to obtain bounds on the conformal
data, see \cite{Poland:2018epd} and references therein for a review.

Let us begin by reviewing crossing symmetry in the case of bosonic blocks. First, in
the $t$-channel, obtained by exchanging $(1\leftrightarrow3)$, one can see that
the cross-ratios \eqref{crossRatios} are exchanged, $u\leftrightarrow v$, and
that the kinematic prefactor gets rescaled by powers of the external dimensions:
\begin{equation}
	K_4 \xrightarrow{1\leftrightarrow3} u^{\frac{\Delta_1+\Delta_2}{2}}v^{-\frac{\Delta_2+\Delta_3}{2}}K_4\,.
\end{equation}
In the case of coincident external bosonic operators with the same conformal
dimension, $\Delta_\phi$, crossing symmetry leads to the well-known sum rule
function $\mathcal{F}_{\Delta,\ell}(u,v)=v^{\Delta_\phi}
g_{\Delta,\ell}(u,v)-u^{\Delta_\phi} g_{\Delta,\ell}(v,u)$, omnipresent in
bootstrap applications. 

Considering the exchange $1\leftrightarrow2$ is also important, as it relates
to the $u$-channel and can lead to important constraints. In that case, the
cross-ratios change according to $(u,v)\leftrightarrow(u/v,1/v)$, and it can be
shown that the bosonic blocks themselves satisfy the relation
\cite{Dolan:2000ut,Dolan:2011dv},
\begin{equation}\label{bosonicExchange12}
	g^{\Delta_{12},\Delta_{34}}(u,v) = (-1)^\ell v^{-\frac{\Delta_{34}}{2}}\, g^{-\Delta_{12},\Delta_{34}}_{\Delta,\ell}(u/v,1/v) \,,
\end{equation}
which in the coincident, non-supersymmetric case acts as a selection rule for
the allowed operators, where only multiplets with even $\ell$
may be exchanged.

While we have used methods involving explicit $SU(2)_R$ indices and auxiliary
variables, $Y^\alpha$, throughout this work we will focus here on crossing
symmetry constraints involving auxiliary variables. This has the distinct
advantage of making the expressions simpler, with all the difficulty hidden in
the details of the $SU(2)_R$ harmonics. The other case leads to equivalent
results, but involves so-called ``Fierz'', or flavour matrices
\cite{Rattazzi:2010yc}. To simplify expressions, we set without loss of
generality the coefficients $a_1=J_1+J_2$\,, $a_2=J_{34}$, and take an ordering such
that $J_1\leq J_2\leq J_3\leq J_4$. Of course, as the full four-point function
does not depend on $a_1$, $a_2$, the choice of convention does not matter in
the end.

Now, under exchange $(1\to3)$, the $SU(2)_R$ cross-ratio is inverted,
$w\to1/w$. Crossing symmetry between $s$- and $t$-channels then imposes the
relation
\begin{equation}\label{stCrossing}
	v^{\frac{\Delta_2+\Delta_3}{2}}\mathcal{G}_\chi(u,v;w) =
	(-1)^{J_1+J_2+J_{34}}w^{-(J_1+J_2)}(1+w)^{J_1-J_3}u^{\frac{\Delta_1+\Delta_2}{2}}\mathcal{G}_\chi(v,u;\frac{1}{w})\,,
\end{equation}
The simplest strategy to obtain the crossing relations satisfied by the
superconformal blocks, $\mathcal{G}_\chi(u,v;w)$, is to remember that the
harmonics are polynomials in $w^{-1}$ of degree $J_1+J_2$, and therefore so are
the blocks. Matching both sides of \eqref{stCrossing} for given superconformal
multiplet order by order in $w^{-1}$, one obtains a certain number of relations and 
supersymmetric versions of $\mathcal{F}_{\Delta,\ell}$ can be defined in a form
suitable for the bootstrap. For coincident R-charges, $J_{i}=J$, there are
$J+1$ such relations, and one can use the Ward identity to find $J$ independent
crossing equations \cite{Chang:2017xmr}.  We note that in the bootstrap, using
dependent crossing equations has been known to lead to numerical instabilities,
see e.g. \cite{Chester:2014fya}.

Let us finish by checking the kind of constraints one obtains when considering the
exchange $(1\leftrightarrow2)$. The $SU(2)_R$ cross ratio changing
according to $w\to -w/(1+w)$, and using the standard Kummer relations for
hypergeometric functions, one obtains
\begin{equation}
	\mathcal{P}_J^{-J_{12},J_{34}}\left (\frac{-w}{1+w} \right) = (-1)^{J}(1+w)^{2J_{34}}\,\mathcal{P}_J^{J_{12},J_{34}}(w)\,.
\end{equation}
Together with the transformations of $K_4,\,K_4^R$, and the bosonic blocks under the same
exchange \eqref{bosonicExchange12}, this gives additional constraints on the
blocks. For instance, in the case where $J_{ij}=0$, this leads to the
selection rule
\begin{equation}\label{constrainCoincident}
	\ell+J_R\in2\mathbb{Z}\,,\qquad\text{if }J_{12}=0=J_{34}\,.
\end{equation}
Notice that the same constraints also apply to descendants,
$\tilde{\ell}+J\in2\mathbb{Z}$. As we will see when solving the constraints
discussed in this section, the superconformal block knows about it, and the
coefficient of a would-be violating descendant precisely vanishes in that case.

\section{Superconformal Blocks of Mixed $\mathcal{D}$-Type Four-Point Functions}\label{sec:Results}

We are now ready to solve the constraints derived in the last section, and find
the explicit form of the superconformal blocks associated to each of the
superconformal multiplets. We consider blocks appearing in the
four-point function of four scalar superconformal primaries, $\phi_i$, belonging 
to short superconformal multiplets, $\mathcal{D}[J_i]$. Finding an explicit
form for the superconformal block involves determining the coefficients,
$f^J_{\tilde{\Delta},\tilde{\ell}}\;$, of the bosonic conformal blocks in the
expansion \eqref{blockDecomposition}. These are determined by considering the
type (II) Casimir equation \eqref{superCasimirEquationTypeII} and the
superconformal Ward identity \eqref{WardIdentity}. As we will find below, while
the type (II) constraint fixes the coefficients,
$f^J_{\tilde{\Delta},\tilde{\ell}}$, for all of the contributing $\mathcal{D}$-type blocks, it
is in general \textit{not} strong enough to find unique solutions for all types
of superconformal multiplets. On the other hand, the Ward identity will be able
to fix uniquely---or forbid---all possible superconformal blocks.

In the general case, where we consider the superconformal blocks associated to
e.g. long multiplets, the expressions for the coefficients are given as
ungainly rational functions of the various conformal data. For the convenience
of the reader who might want to use them, they have been included in a
\texttt{Mathematica} file attached to the arXiv submission, and they are
directed there for the complete list of superconformal blocks. Here, we write
the results for some of the edge cases, such as the superconformal blocks for
the short multiplets, or some of the more manageable coefficients of the
long superconformal multiplets.

\subsection*{Solving the Constraints}\label{sec:solvingConstraints}
As mentioned already, the constraints follow from applying differential
operators to the bosonic blocks. With the exception of
$\mathcal{D}_\text{bos}$, the bosonic blocks are not
eigenfunctions of these operators, and this must be dealt with. To do so, we use
introduce a decomposition in terms of Jack polynomials, $P^{(\varepsilon)}_{\lambda_1,\lambda_2}$, 
of the bosonic blocks\cite{Dolan:2003hv,Dolan:2011dv},
\begin{equation}
	g^{\Delta_{12}, \Delta_{34}}_{\Delta, \ell}(z,\bar{z}) = \sum_{m,n \geq0} r_{m,n}(\Delta_{12}, \Delta_{34}, \Delta, \ell) P^{(\varepsilon)}_{\frac{1}{2}(\Delta + \ell) + m, \frac{1}{2}(\Delta - \ell) + n}(z,\bar{z})\,.
\end{equation}
The coefficients, $r_{m,n}$, of the infinite series depend solely on the
conformal data, and can be determined recursively by application of the bosonic
Casimir operator \eqref{bosonicCasimirEquation}, 
\begin{align}\label{rmn}
	r_{m,n}({\Delta_{12}, \Delta_{34}, \Delta, \ell}) = &
	\left( \frac{1}{2}( \Delta + \ell - \Delta_{12} ) \right)_m
	\left( \frac{1}{2} ( \Delta + \ell + \Delta_{34}) \right)_m \times\nonumber\\
	&\left( \frac{1}{2} ( \Delta - \ell - \Delta_{12}) - \varepsilon \right)_n
	\left( \frac{1}{2} ( \Delta - \ell + \Delta_{34} ) - \varepsilon  \right)_n 
	\widehat{r}_{m,n}^{\Delta, \ell}\,,
\end{align}
where $(a)_{n}$ denotes the Pochhammer symbol. The quantity $\widehat
r_{m,n}^{\Delta, \ell}$ can be evaluated recursively via the relation 
\begin{align}\label{rHatmn}
	(m(m+\Delta + \ell - 1 ) &+ n( n+\Delta - \ell - 2 \varepsilon - 1)) \,\,  \widehat r_{m,n}^{\Delta, \ell}\nonumber\\
	&= \frac{ \ell + m - n - 1 + 2 \varepsilon }{ \ell + m - n - 1+ \varepsilon} \widehat{r}_{m-1,n}^{\Delta, \ell}
	+ \frac{\ell + m - n +1}{\ell + m - n + 1 + \varepsilon } \widehat r_{m,n- 1}^{\Delta, \ell}\,,
\end{align}
together with the initial condition $r_{0,0} = 1$. We note that a closed form
of this recursion relation is known in terms of a generalised hypergeometric
function, ${}_4F_3$, but it will not be necessary here \cite{Dolan:2003hv}. 

Using the properties of the Jack polynomials collected in appendix
\ref{app:superconformalcasimir}, one can infer the action of
$\mathcal{D}_\text{SUSY}$ as 
\begin{align}
	\mathcal{D}_\text{SUSY}P^{(\varepsilon)}_{\lambda_1, \lambda_2} &= 
	4 (\lambda_1 + \lambda_2) P^{(\varepsilon)}_{\lambda_1, \lambda_2} 
	-2 \frac{(\Delta_{34}+2 \lambda_1) (\lambda_1-\lambda_2+2 \varepsilon ) }{\lambda_1-\lambda_2+\varepsilon } P^{(\varepsilon)}_{\lambda_1+1,\lambda_2} \nonumber\\
	&~ ~ -2 \frac{(\lambda_1-\lambda_2) (\Delta_{34}+2 \lambda_2-2 \varepsilon )}{\lambda_1-\lambda_2+\varepsilon }  P^{(\varepsilon)}_{\lambda_1,\lambda_2+1}\, .
\end{align}
This formula has no explicit dependence on $z,\bar{z}$ and makes the Jack
polynomials very good candidates to deal with the full superconformal Casimir
operator. Indeed, the action of the Casimir operator on a conformal block gives
a linear combination of Jack polynomials. As there is no explicit dependence
on spacetime coordinates, one can use the fact that they form an orthogonal
basis of symmetric polynomials to solve the constraints order by order. This
method has proven very efficient to find the form of the superconformal blocks
of momentum map operators in this context, see e.g. \cite{Bobev:2015jxa,Bobev:2017jhk}. 

In the case of the Ward identity, we have not determined a way to write the
differential operator \eqref{WardIdentity} on a Jack polynomial in a closed
form, i.e. in such a way that it involves only a linear combination of Jack
polynomials without explicit dependence on the coordinates. In
\cite{Chang:2017xmr}, the Ward identity associated to external momentum maps
operators, $\mathcal{D}[1]$, was solved using various inversion formulae
\cite{Dolan:2004mu}. While these formulae can be generalised for higher
external spins, we find the following procedure more convenient: use the
properties of Jack polynomials (see appendix \ref{app:superconformalcasimir})
and of hypergeometric functions to reduce the Ward identity to a manageable,
albeit long, form; perform an expansion in radial coordinates, in the same
spirit of \cite{Hogervorst:2013sma}; and finally solve the constraints order by
order.

In all cases, the above procedures lead to a linear set of equations involving
the superconformal block coefficients of the form
\begin{equation}\label{LinearSystem}
	\sum_{J,\tilde{\Delta},\tilde{\ell}} \alpha_{\tilde{\Delta},\tilde{\ell}}^J\; f^{J}_{\tilde{\Delta},\tilde{\ell}} = 0\,,
\end{equation}
where the sum is taken over all possible non-supersymmetric conformal primaries
inside the multiplet under consideration, and
$\alpha_{\tilde{\Delta},\tilde{\ell}}^J$ is a function of the superconformal
data, possibly vanishing. It is then clear that one of two possibilities can
occur: either all the coefficients are proportional to that of the
superconformal primary---that we safely set to one---or they all vanish and the
multiplet does not appear in the block expansion. 
While tedious, determining the set of constraints from the above procedure can
be implemented algorithmically in a straightforward manner, up to a given
order.
If a solution is found to be valid up to
some threshold, one can then easily check that it is satisfied to
all orders by reinjecting it in the original equation and using the recursion
relations \eqref{rmn} and \eqref{rHatmn}.

Let us now outline the result of this procedure and comment on some features for
all types of superconformal multiplets, going from simplest to most
complex (i.e. largest size of the) multiplets.

\subsubsection*{$\mathcal{D}$-Type Multiplets}

Let us apply both methods to the simplest class of multiplets, type
$\mathcal{D}[J_R]$. In this case, the terms in the expansion
\eqref{blockDecomposition} of the superconformal block are particularly simple,
making them a good stepping stone to the other, more complicated, multiplets.
Indeed, their primaries can only be scalars, and there are only three possible
states in the $\ell$-symmetric representation of the Lorentz group:
\begin{equation}
	(\tilde{\Delta},\tilde{\ell},J) \in \left\{ (2\varepsilon J_R,0,J_R)\,,(2\varepsilon J_R+1,1,J_R-1)\,,(2\varepsilon J_R+2,0,J_R-2) \right\}\,.
\end{equation}
The first entry of the list corresponds to the superconformal primary.
Plugging in this spectrum in both the Casimir and Ward identity equations, all
constraints can be uniquely solved and the following values for the
coefficients of the superconformal blocks are found: 
\begin{align}
	f_{2 \epsilon  J_R+1,1}^{J_R-1} &=  \frac{\epsilon (J_R+J_{12})(J_R+J_{34})}{J_R (2 \epsilon J_R+1)}\,,\\
f_{2 \epsilon  J_R+2,0}^{J_R-2} &= \frac{\epsilon^2(J_R+J_{12})(J_R+J_{34})(J_R+J_{12}-1)(J_R+J_{34}-1) }{2 (J_R-1) (2 J_R-1) (2 \epsilon  J_R+1) (2 \epsilon  J_R-\epsilon +1)}\,.
\end{align}
These results generalise those found in four dimensions,
\cite{Nirschl:2004pa,Beem:2014zpa}, and for external momentum maps
\cite{Chang:2017xmr,Bobev:2017jhk}. 

Note that the structure of the blocks naturally encodes a lot of information
about the representation theory of the conformal group.  For instance, the
trivial multiplet, $\mathcal{D}[0]$, corresponds to the identity operator,
which has no superpartners and can only appear if $J_{12}=0=J_{34}$. In that
case there is only one coefficient in the expansion of the superconformal
block, that of the lone primary, while the others vanish.  Another example is
that of the momentum map, $\mathcal{D}[1]$, containing a null-state at level
$J_R-2$ corresponding to the conservation of the flavour current,
$\partial_{\mu\,}\mathcal{J}^\mu=0$. When it is allowed in the OPE, the block
``knows'' that there is no corresponding state at that level, and its
coefficient vanishes. The same kind of remarks will also apply with different
conserved quantities in other types of multiplets.

We are left with one constraint still unchecked, that of the type (I) Casimir
equation \eqref{typeIConstraint}. For $\mathcal{D}$-type multiplets, a quick
inspection of all possibilities reveals that the only state satisfying the
constraint is the primary. This means that any such multiplet with R-charge
$J_R\in \mathcal{J}$ is allowed as an exchanged operator in the decomposition. 

\subsubsection*{$\mathcal{B}$-type Multiplets}

For $\mathcal{B}$-type multiplets, the states that can possibly appear in the superconformal blocks are 
\begin{align}\label{BTypeSpectrum}
	(\tilde{\Delta}, \tilde{\ell}, J)\in\big\{&
		(\Delta, \ell, J_R)\,,\nonumber\\
		&(\Delta+1, \ell\pm1, J_R-1)\,,
		(\Delta+1, \ell+1, J_R)\,,
		(\Delta+1, \ell+1, J_R+1)\,,\nonumber\\
		&(\Delta+2, \ell, J_R-2)\,,
		(\Delta+2, \ell, J_R-1)\,,
		(\Delta+2, \ell, J_R)\,,
		(\Delta+2, \ell+2, J_R)\,,\nonumber\\
		&(\Delta+3, \ell+1, J_R-1)
	\big\}\,.
\end{align}

In this case, the type (I) Casimir equation is never satisfied by the primary,
which excludes $\mathcal{B}[\ell,J_R=J_\text{max}]$ to appear in the block
decomposition. The state with $(J_R+1)$ is however the only one satisfying the
constraints, which mean that the biggest R-charge these multiplets can
achieve is $J_R=J_\text{max}-1$.

The type (II) Casimir equation is no longer strong enough to fix all of the
coefficients due to an increased number of possible states. The linear system
\eqref{LinearSystem} is indeed underdetermined and one cannot find a unique
solution. Previous works \cite{Bobev:2015jxa,Bobev:2017jhk} looked at cases
where the type (II) Casimir equation was always enough to get a unique
solution, but this fails when considering more involved cases. This can be
disheartening, but we have a second set of constraints: the Ward identity.
Such an approach does not suffer from this drawback and is in fact able to find
a unique solution for any multiplet. One can check that the underdetermined
system of the Casimir is satisfied by that solution. Due to the lengthy
expression of the results, we have tabulated them in appendix
\eqref{app:resultCoefficients}. 

Note that in section \ref{sec:CrossingSymmetry}, we have seen that if all
external scalars have the same R-charge, the states in equation
\eqref{BTypeSpectrum} must satisfy $J+\tilde{\ell}\in2\mathbb{Z}$. There are
two states violating this condition: $(\Delta,\ell+1,J_R)$ and $(\Delta,
\ell,J_R-1)$. As one can see from equation \eqref{ResultBType}, the
superconformal blocks know about this selection rule and their coefficients
automatically vanish should the primary be allowed.

\subsubsection*{$\mathcal{C}$- and $\mathcal{A}$-Type Multiplets}

As has been already mentioned in the introduction, the $\mathcal{C}$-type is
special to six dimensions due to the presence of tensor multiplets. In that
case, both the Casimir equation and the Ward identity lead to solutions only if
all coefficients, including that of the primary, vanish. This indicates
that the superconformal blocks for these multiplets cannot appear in the block
decomposition of the four-point function of any half-BPS states. Again, this
generalises the results of momentum map operators to higher external
R-charges \cite{Ferrara:2001uj,Chang:2017xmr}.

We find a similar behaviour in the case of type $\mathcal{A}$. While the
Casimir equation again fails to fix all coefficients uniquely, the Ward
identity ultimately shows that the superconformal block is trivial. From this,
we conclude that both $\mathcal{C}$- and $\mathcal{A}$-type multiplets do not
appear in the block expansion. 

\subsubsection*{Long Superconformal Multiplets}

Finally, we arrive at a discussion of the long multiplets. Such multiplets
contain a total of twenty-one different states in a traceless-symmetric
representation of the Poincar\'e group, which we summarise in table
\ref{tab:LongSpectrum}. An analysis of these states against the type (I)
Casimir equation reveals that none of them can satisfy it, except for the lone
state with R-charge $J=J_R+2$, which implies that long multiplets can only
appear in the decomposition if they have $J_R\leq J_\text{max}-2$, as was
claimed in equation \eqref{blockDecomposition}.

The type (II) Casimir equation is again not able to find a unique solution for
the full superconformal block, while the Ward identity fixes it completely. The
procedure for long multiplets remains completely the same as for short multiplets, albeit
for much longer expressions due to the fact that conformal dimensions are not
fixed by unitarity.
\begin{table}
	\centering
	\begin{tabular}{|c|c|}\hline
		$J_R-2$ & $(\Delta +2,\ell)$\\\hline
		$J_R-1$ & $(\Delta +1,\ell\pm1)\,,(\Delta +2,\ell)\,,(\Delta +3,\ell\pm1)$\\\hline
		$J_R$   & $(\Delta, \ell)\,,(\Delta+1,\ell\pm1)\,,(\Delta+2,\ell\pm2)\,,(\Delta+2,\ell)\,,(\Delta+3,\ell\pm1)\,,(\Delta+4,\ell)$\\\hline
		$J_R+1$ & $(\Delta +1,\ell\pm1)\,,(\Delta +2,\ell)\,,(\Delta +3,\ell\pm1)$\\\hline
		$J_R+2$ & $(\Delta +2,\ell)$\\\hline
	\end{tabular}
	\caption{$\ell$-symmetric states potentially appearing in a long superconformal multiplet.}
	\label{tab:LongSpectrum}
\end{table}
The results, for all the allowed coefficients,
$f^J_{\tilde{\Delta},\tilde{\ell}}$, quickly become long and unwieldy, and as such we
have included an exhaustive list of the coefficients in a \texttt{Mathematica}
notebook attached to the arXiv submission of this article.
We however have written a selected list of the simplest coefficients in appendix
\ref{app:resultCoefficients}. 

Interestingly, we find that---at least some of---the coefficients
are related to one another by linear shifts of the conformal data. For
instance, if two states $(\tilde{\Delta},\tilde{\ell},J\pm j)$ both appear in the
spectrum for a given $j$, they are related to one another by a linear
transformation of $J_R$ in their respective expression. By looking at level one
for instance, equation \eqref{longMultipletLevel1}, it is easy to convince
oneself that
\begin{align}
	f_{\Delta+1, \ell\pm1}^{J_R-1} =& \left. f_{\Delta+1, \ell\pm1}^{J_R+1}\right|_{J_R\to -(J_R+1)}\,.
\end{align}
This behaviour between fixed conformal parameter and R-charge generalises to
higher levels, sometimes involving shifts of $\ell$ as well
\begin{align}
	f_{\Delta+2, \ell}^{J_R-2} =& \left. f_{\Delta+2, \ell}^{J_R+2}\right|_{J_R\to -(J_R+1)}\,,\nonumber\\
	f_{\Delta+2, \ell}^{J_R-1} =& \left. f_{\Delta+2, \ell}^{J_R+1}\right|_{J_R\to -(J_R+1),\ell\to-(\ell+2\varepsilon)}\,,\\
	f_{\Delta+3, \ell}^{J_R-1} =& \left. f_{\Delta+2, \ell}^{J_R+1}\right|_{J_R\to -(J_R+1)}\,.\nonumber
\end{align}
Similar relations exist for fixed R-charge and shifted Poincar\'e
representation:
\begin{align}
	f^{J}_{\tilde{\Delta},\ell+1} =& \left.(f^{J}_{\tilde{\Delta},\ell-1})\right|_{\ell\to-(\ell+2\varepsilon)}\,.
\end{align}
This relation is valid for all fixed $(\tilde{\Delta},J)$ with $\ell\pm1$
appearing in table \ref{tab:LongSpectrum}. Those linear transformations can be
traced back to the four accidental $\mathbb{Z}_2$ symmetries of the
superconformal Casimir \eqref{supersymmetricCasimir},
\begin{align}
	\big(\Delta,\ell\big)&\longleftrightarrow \big(-(\ell+1),-(\Delta+1)\big)\\
	\Delta&\longleftrightarrow -(\Delta+D-4)\\
	\ell&\longleftrightarrow -(\ell+(D-2))\\
	J_R&\longleftrightarrow -(J_R+1)
\end{align}

While we were unable to find relations between states that are not shifted from
the data of the primary by the same increment, e.g. finding a transformation between
$f^{J_R}_{\tilde{\Delta},\tilde{\ell}}$ and $f^{J_R-1}_{\tilde{\Delta},\tilde{\ell}}$ or the more
complicated coefficients, they hint at a perhaps more fundamental form of the
superconformal blocks. It was indeed pointed out in \cite{Fitzpatrick:2014oza,
Bobev:2015jxa} that in four dimensions, the superconformal Casimir equation for
coincident external primaries in $\mathcal{N}$-extended supersymmetry is
satisfied by
$\mathcal{G}=u^{-\mathcal{N}/2}g_{\Delta+\mathcal{N},\ell}^{\Delta_{12}=\Delta_{34}=\mathcal{N}}(u,v)$,
although it is not obvious that such that such an expression can be decomposed
back into bosonic blocks with $\Delta_{ij}=0$.  Our findings hint at a
generalisation of this expression to arbitrary dimensions.

Finding such a fundamental form of the superconformal blocks could potentially
enable to find faster algorithms for the superconformal bootstrap, leading to
stricter bounds, as the form above is considerably simpler than that of
equations \eqref{BlockUncontracted} or \eqref{superconformalblock} with the coefficients as determined herein. We leave
further analysis in this direction for future work.

We close this section by commenting on the blocks of long multiplets in the
limit when the external fields have the same R-charge. While this certainly
makes the expressions simpler, the coefficient $f_{\Delta+2,\ell}^{J_R}$
remains
particularly long. A simplification of note is that there are six coefficients
that
vanish identically in that limit. As in the case of $\mathcal{B}$-type
multiplets, this is traced back to the condition \eqref{constrainCoincident}
that these states must satisfy $\tilde{\ell}+J=0\mod 2$.

\section{Conclusions}\label{sec:Conclusions}

In this work we provide explicit expressions for the superconformal blocks
associated to each superconformal multiplet that appears inside of the four-point
function of scalars contained inside of $\mathcal{D}$-type multiplets for
theories with eight supercharges in arbitrary dimensions $2<d\leq6$. These
results generalise previous work
\cite{Chester:2014mea,Liendo:2016ymz,Beem:2014zpa,Lemos:2015awa,Chang:2017cdx,Chang:2017xmr,Bobev:2017jhk}
on the topic to arbitrary R-symmetry representations and to mixed correlators
of primaries from different $\mathcal{D}$-type multiplets. In particular, we find
that also for $R$-spin greater than one, only three out of the five types of
superconformal multiplets are allowed in the superconformal block expansion of
the four-point function of \emph{a priori} distinct superconformal primaries of
half-BPS multiplets.

The Ward identity was, as expected, powerful enough to find a unique solution
for all the blocks, where such a computation was feasible to do, or else to
exclude them from the expansion altogether. On the other hand the Casimir equation was not able
to fix the form of the superconformal blocks completely, but only impose a set
of relations between the involved coefficients. One could think that other
constraints might come from the richer structure of the projectors that was
not present in previous works, but even for coincident external primaries,
where the form of the type (II) Casimir equation is unique, we were not able
to determine all coefficients. It may be the case that in order to find a
unique solution, one must resort to also studying the Casimir equations that
arise when considering correlation functions that do not involve states
carrying the extremal weights of the $SU(2)_R$ representations, i.e., to go
beyond the conditions discussed in 
\eqref{halfBPSCondition}. One would indeed expect that the Casimir equation
contains the same amount of information about the blocks as the Ward identity.

These solutions follow an intriguing web of transformation relating them,
leading one to think that there might be a more fundamental structure of the
superconformal blocks. It would be interesting to see if this is related
to---or at least could help understand---the algebraic structure of protected
operators discovered in four-dimensional $\mathcal{N}=2$ SCFTs and its
generalisations \cite{Beem:2013sza}. Finding a more compact form for the blocks
could also potentially improve the rapidity of current numerical applications to the
superconformal bootstrap.

The most obvious application of our results is indeed to the numerical
bootstrap. So far, most of the literature has focused on external primaries
associated to half-BPS multiplets containing flavour currents or the
energy-momentum tensor, depending on the number of supercharges. Our results
enable a line of research going beyond this and potentially allows for an
extension of the known set of
bounds. Furthermore we have not made any special kind of assumption for
the nature of the external primaries except that they are of type
$\mathcal{D}[J_i]$. The blocks also enable a study of mixed correlators, which
have led to strong bounds on conformal data and insights on possibly minimal
supersymmetric CFTs
\cite{Kos:2014bka,Kos:2015mba,Lemos:2015awa,Kos:2016ysd,Li:2016wdp,Li:2017ddj
,Rong:2018okz,Kousvos:2018rhl,Agmon:2019imm}.

Due to their independence of the dimension of spacetime, our results are also
predisposed to studies on six-dimensional SCFTs, a field that has remained
quite unexplored compared to lower dimensions (see
\cite{Beem:2015aoa,Chang:2017xmr} for the works that initiated the program).
Furthermore a conjecturally complete classification of these theories has been
achieved using string theory
\cite{Heckman:2013pva,Heckman:2015bfa,Bhardwaj:2018jgp} (see
\cite{Heckman:2018jxk} for a review). The relationship between conformal field
theory and the compactification geometry seems to indicate that the conformal
data might be tightly controlled by geometric invariants. As an example, the
central charge of these theories is related to intersection numbers of the
underlying elliptic fibration via anomaly polynomial relations. A natural
question is then to ask whether the bootstrap can shed some light on the
nature of these invariants; this topic is the subject of work that will appear
in the near future \cite{nextPaper}.


\subsection*{Acknowledgements}

\noindent We thank F. Apruzzi, M. Gillioz, and M. Wiesner for helpful
discussions. F.B. is especially grateful to Riccardo Rattazzi and the theory
group at EPFL for hospitality and stimulating discussions in the infancy of
this project. C.L. would like to thank the 2019 Pollica Summer Workshop,
supported in part by the Simons Collaboration on the Non-perturbative
Bootstrap and in part by the INFN, for a stimulating environment during
important stages of this work.  The work of F.B.  and M.F. is supported
through the grants SEV-2016-0597, FPA2015-65480-P and PGC2018-095976-B-C21
from MCIU/AEI/FEDER, UE. C.L. is supported by NSF CAREER grant PHY-1756996.


\appendix 
\section{The Superconformal Group}\label{app:superconformalGroup}

We are interested in superconformal field theories whose bosonic subalgebra is
$\mathfrak{so}(1,d+1)\times \mathfrak{su}(2)_R$. We follow the work of
\cite{Minwalla:1997ka} and impose a reality constraint on the generators
compatible with unitarity in Lorentzian signature.  Namely, we choose a basis of
generators satisfying the constraints
\begin{equation}
	D^\dagger = -D\,,\qquad 
	M_{\mu\nu}^\dagger = M_{\mu\nu}\,,\qquad 
	P_\mu^\dagger = K_\mu\,,\qquad
	R_i^\dagger = R_i\,,
\end{equation}
where: $(P_{\mu}\,,M_{\mu\nu})$ are the usual generators of the Poincar\'e
group; $D$ and $K$ are the generators of dilatations and special conformal
transformations respectively; and $R_i$ the generators of the R-symmetry group.
Chosen so, the generators satisfy the usual Euclidean conformal algebra
\begin{gather}
	[M_{\mu\nu},M_{\rho\sigma}] = -i\left( \delta_{\mu\sigma}M_{\nu\rho}+\delta_{\nu\rho}M_{\mu\sigma}-\delta_{\mu\rho}M_{\nu\sigma}-\delta_{\nu\sigma}M_{\mu\rho} \right)\,,\\
	[M_{\mu\nu},P_\rho] = -2iP_{[\mu}\delta_{\nu]\rho}\,,\qquad
	[M_{\mu\nu},K_\rho] = -2iK_{[\mu}\delta_{\nu]\rho}\,,\\
	[D,P_\mu] = -iP_\mu\,,\qquad
	[D,K_\mu] = iK_\mu\,,\\
	[P_\mu,K_\nu] = -2i(\delta_{\mu\nu}D+M_{\mu\nu})\,,\qquad
	[R_i,R_j] = i\varepsilon_{ijk}R_k\,.
\end{gather}
In addition to the bosonic charges, there are two sets of fermionic
generators, $Q_{\alpha A}\,, S^{\alpha A}$. They satisfy the conjugation
rule, $S^{\alpha A} = (Q_{\alpha A})^\dagger$. The anti-commutation relations
satisfied by these fermionic generators can be obtained by considering the
six-dimensional SUSY algebra and reducing to lower dimension. It was shown in
\cite{Bobev:2017jhk} that to be closed under Jacobi identities, they must
take the form :
\begin{gather}
	\left\{ Q_{\alpha A},Q_{\beta B}  \right\}= \varepsilon_{\alpha \beta}\Gamma_{AB}^\mu P_\mu\,,\qquad
	\left\{ S^{\alpha A},S^{\beta B}  \right\}= \varepsilon_{\alpha \beta}\widetilde{\Gamma}^{AB}_\mu K_{\mu}\,,\\
	\left\{ S^{\alpha A},Q_{\beta B}  \right\}= i\delta^\alpha _\beta \delta^A_B D - (d-2)\delta^A_B R^\alpha _\beta+ \delta_\beta^\alpha  (m_{\mu\nu})^A_B M_{ij} \,,
\end{gather}
with $(m_{\mu\nu})=- \frac{i}{2}\widetilde{\Gamma}_{[\mu}\Gamma_{\nu]}$ and
$R^\alpha_\beta=R_i(\sigma_i)^\alpha_\beta$, where $\sigma_i$ are the Pauli
matrices. As we are considering theories with eight supercharges, $N_S=8$,
one has $\delta_\alpha^\alpha\delta^A_A=8$. This can be generalised to a
higher or lower number of supercharges by changing
$\varepsilon_{\alpha\beta}\to\Omega_{\alpha\beta}$ to the appropriate
pairing. Moreover, the precise form of $\Gamma^\mu\,,\widetilde{\Gamma}_\mu$
depends on the considered dimension, but it can be shown that they satisfy a
Clifford-like identity,
$\widetilde{\Gamma}_{(\mu}\Gamma_{\nu)}=\delta_{\mu\nu}\mathbf{1}$. This
relation will be sufficient in this work, and we will not have to deal with
particular explicit expressions of the spinor matrices.

The mixed fermionic-bosonic identities can be determined via Jacobi identities:
\begin{gather}
	[K_\mu, Q_{\alpha A}] =  \varepsilon_{\alpha \beta} \Gamma^{\mu}_{AB} S^{\beta B}\,,\qquad
	[P_\mu, S^{\alpha A}] = -\varepsilon^{\alpha \beta} \widetilde{\Gamma}_{\mu}^{AB} Q_{\beta B}\,,\\
	[M_{\mu\nu},Q_{\alpha A}] =  (m_{\mu\nu})^{B}_{\phantom{B}A}Q_{\alpha B}\,,\qquad
	[M_{\mu\nu},S^{\alpha A}] = -(m_{\mu\nu})^{A}_{\phantom{A}B}S^{\alpha B}\,,\\
	[R_i ,Q_{\alpha A}] =  \frac{1}{2}(\sigma_i)_\alpha^\beta Q_{\beta A}\,,\qquad
	[R_i ,S^{\alpha A}] = -\frac{1}{2}(\sigma_i)^\alpha_\beta S^{\beta A}\,,
\end{gather}

The quadratic Casimir is then given by\footnote{ In order to make the Casimir
and $\left\{ S^{\alpha A},Q_{\beta B} \right\}$                      dimension
independent, one needs to add contributions from the transverse
coordinates $d<\hat{\mu}\leq 6$. As we are only interested in
traceless-symmetric representations of the Lorentz group in a given
dimension, and therefore uncharged under extra R-symmetry factors in this
work, we will not take these additional pieces into account. We refer to
\cite{Bobev:2017jhk} for more details.}
\begin{align}
	C^2  &= C_\text{bos}^2 + C_\text{SUSY}^2 + C_R^2\,,\\
	C_\text{bos}^2 &= \frac{1}{2} M_{\mu\nu}M^{\mu\nu}-D^2 - P_{(\mu}K_{\nu)}\,,\\
	C_\text{SUSY}^2 &= \frac{1}{2}[S^{\alpha A},Q_{\alpha A}]\,,\qquad C_R^2 =  - 2\varepsilon R_iR^i\,.
\end{align}

\section{Conformal Frame and $SU(2)_R$ Harmonics}\label{app:ConformalFrame}

When dealing with homogeneous function such as the four-point function, it is often
useful to go to a specific conformal frame, where the dependence on given
coordinates is clearer. In this work, we deal with three different sets of
coordinates, related to the conformal cross-ratios \eqref{crossRatios}:
\begin{equation}\label{RsymmetryFrame2}
	u = z \bar{z} = \frac{ \chi \bar{\chi} }{(1 + \chi) ( 1+ \bar{\chi})} \, , \qquad 
	v = (1+ z)(1+ \bar{z}) = \frac{1}{(1 + \chi) ( 1+ \bar{\chi})}\,.
\end{equation}
The coordinates $u,v$ and $z,\bar{z}$ are by now standard in the
literature, see e.g.
\cite{Rychkov:2016iqz,Simmons-Duffin:2016gjk,Poland:2018epd} and references
therein. The third, $\chi$, can be obtained by going in the following Lorentzian frame. 
\begin{align}
	x_1 &= \left( 1+ {1 \over 2} (\chi + \ov \chi), {1 \over 2} ( \chi - \ov \chi) , 0 , \dots, 0  \right)\,\\
	x_2 &= (1, 0 , \dots , 0), \qquad x_3 = (x_4)^{-1} = 0\,.
\end{align}
The kinematic prefactor, $K_4$ in the four-point function then turns into 
\begin{equation}
	K_4 \rightarrow(\chi \ov \chi)^{\frac{-(\Delta_1+\Delta_2)}{2}  } ((\chi+1) (\bar{\chi}+1))^{\frac{ -\Delta _{34} }{2}}\, ,
\end{equation}
where $\Delta_{ij} = \Delta_i - \Delta_j$. These coordinates are quite useful,
as explained in the main text, as under a supersymmetric transformation in
superspace, they acquire a pole
\begin{equation}
	\delta \chi \sim \frac{1}{w-\chi}  \, , \qquad 
	\delta w = \varepsilon\, \delta \chi\,.
\end{equation}

\subsection*{$SU(2)_R$ Harmonics}
We now derive the $SU(2)_R$ harmonics, \eqref{hyperGeometricHarmonic}, by inserting the
quadratic Casimir of $SU(2)_R$ into the four-point function. Let us start with
the four-point function of four---in general non-identical---primaries depending on
the auxiliary variable, $Y^\alpha$ introduced in section
\ref{sec:RsymmetryVariables},
\begin{equation}\label{app:fourPointFunction}
	\left< \phi_1(x_1, Y_1)\phi_2(x_2, Y_2)\phi_3(x_3, Y_3)\phi_4(x_4, Y_4)  \right> =K_4K_4^R F(u,v;w)\, .
\end{equation}
From $SU(2)_R$ invariance it is clear that the four-point function may depend
on the $Y_i$ through the $SU(2)_R$ invariant product 
\begin{equation}
	Y_{ij} := Y_i \cdot Y_j = Y_i^{\alpha} Y_j^\beta \varepsilon_{\alpha \beta} \, .
\end{equation}
Furthermore the primaries, $\phi_i(x_i,Y_i)$, must be homogeneous functions
of degree $(-\Delta_i,2J_i)$, and the correlator
\eqref{app:fourPointFunction} will depend on a single scale invariant cross
ratio, $w$, satisfying a completeness relation:
\begin{equation}
	w = \frac{(Y_1 \cdot Y_2) (Y_3 \cdot Y_4)}{(Y_1 \cdot Y_4) (Y_2 \cdot Y_3) } \,,\qquad
	1+w = \frac{(Y_1 \cdot Y_3) (Y_2 \cdot Y_4)}{(Y_1 \cdot Y_4) (Y_2 \cdot Y_3) } \,.
\end{equation}
The homogeneity property of \eqref{app:fourPointFunction} allows to separate
the four-point function into a prefactor, $K_4^R$, which has the correct
scaling and a function of the invariant cross-ratio, $w$. Enforcing this
scaling behaviour in an ansatz, $K_4^R = Y_{12}^{a_1} Y_{13}^{a_2} Y_{14}^{a_3}
Y_{23}^{a_4}  Y_{24}^{a_5}  Y_{34}^{a_6}$, fixes four out of six parameters,
leading to the most general form,
\begin{equation}
	K_4^R=
	\left(Y_{12}\right){}^{a_1} \left(Y_{13}\right){}^{a_2}
	\left(Y_{14}\right){}^{-a_1-a_2+2 J_1}
	\left(Y_{23}\right){}^{-a_1-a_2+J^+_{12}+J_{34}} 
	\left(Y_{24}\right){}^{a_2-J_{12}-J_{34}} 
	\left(Y_{34}\right){}^{a_1-J^+_{12}+J_{34}^+}\,, 
\end{equation}
where we defined $J_{ij} = J_i - J_j$ and $J^+_{ij} = J_i + J_j$. Notice that
two parameters $a_1,a_2$ are still undetermined, and correspond to rescaling of
the function $F(u,v;w)$ (see equation \eqref{fourpt}) by factors of $w$ and
$(1+w)$. To better see this, it is instructive to go to a specific R-symmetry
frame, which is used in the derivation of the superconformal Ward identity. It
is defined as
\begin{align}\label{RsymmetryFrame}
	Y_i   &= (1 , y_i) ~\Rightarrow~ Y_{ij}  = y_i - y_j\,,\nonumber\\
	y_1   &= 1+ w \, , \qquad y_2 = 1\, , \qquad y_3 = y_4^{-1} = 0\,,\\
	K_4^R &\rightarrow w^{a_1} (1+w)^{a_2}  \, .\nonumber
\end{align}
Therefore different choices of $a_1,\,a_2$ can be taken care of by
absorbing $w^{a_1} (1+w)^{a_2} $ into the definition of $F(u,v;w)$. 

We will now determine the $w$ dependence of $F(u,v;w)$ by determining the
harmonic functions of $SU(2)_R$ by deriving the differential equation
associated to the quadratic Casimir $C^2=R_iR^i$. The logic of the derivation is
well known, but by the simplicity of $SU(2)$ allows to show an easy example
that can then be applied \emph{mutatis mutandis} to the superconformal Casimir
in section \ref{app:superconformalcasimir}. 

The first step of the procedure is to insert a projector
\begin{equation}
	\text{Proj}_{J} = \sum_{m} \left|J,m \right> \left< J,m \right|\,,
\end{equation}
and the R-symmetry Casimir in the middle of the four-point function to obtain
the contribution of single R-symmetry channel, $F_J(u,v;w)$. One has then two
options: let the Casimir act directly on the projector,
\begin{equation}\label{HarmonicsRHS}
	\left< \phi_1 \phi_2 C_2 (\text{Proj}_J) \phi_3 \phi_4 \right> = J(J+1)  \, K_4 \,	K_4^{R} \, F_J(u,v;w) \,,
\end{equation}
or to let it act on $\phi_1\phi_2$---or $\phi_3\phi_4$, leading to the
same results---which is achieved by letting the generators act as linear
differential operators,
\begin{equation}
	\mathcal{D}_i\mathcal{O}(Y) = \frac{1}{2} (\sigma_i)^{\alpha}{}_{\beta} Y^\beta \frac{\partial\mathcal{O}}{\partial Y^\alpha}\,,
\end{equation}
where $\left\{\sigma_i\right\}$ are the Pauli matrices. As the Casimir acts on both $\phi_1$ and
$\phi_2$, the differential operator is finally:
\begin{equation}\label{HarmonicsLHS}
	\left< \phi_1 \phi_2 C_2 (\text{Proj}_J) \phi_3 \phi_4 \right> = 
	(\mathcal{D}_1+\mathcal{D}_2)_i(\mathcal{D}_1+\mathcal{D}_2)^i\, \big(K_4 \,	K_4^{R} \, F_J(u,v;w)\big)\,.
\end{equation}
Comparing equations \eqref{HarmonicsRHS}, \eqref{HarmonicsLHS} and
evaluating the differential operators leads to the second order differential
equation 
\begin{equation}
	a(w) \partial_w^2F_J(u,v;w)+ b(w) \partial_w F_J(u,v;w) + c(w) F_{J}(u,v;w)= 0 \,, 
\end{equation}
where the coefficients are found to be 
\begin{align}
	a(w) &= w^2 (1 + w)\,,\qquad
	b(w) = 2w(1+w)(a_1-J_{12}^+)+w^2\left( 1+2a_2-J_{12}-J_{34} \right)\,,\nonumber\\
	c(w) &= -J (J+1) +\bigg(\frac{w^2}{1+w}a_2 (a_2-(J_{12}+J_{34}))+w J_{12}J_{34}+(1+w)(a_1-J_{12}^+)^2\\
	&\quad+(a_1-J_{12}^+)(w(2 a_2-J_{12}-J_{34})-1)\bigg)\,.\nonumber
\end{align}
This differential equation can be recast into a more familiar form whose
polynomial solution involves a hypergeometric functions:\footnote{Notice that
the differential equation is second order and therefore has another solution,
also in terms of a hypergeometric function, which can in our case be
disregarded as it is non-polynomial.} 
\begin{equation}\label{generalisedLegendre}
	F_J(u,v;w) = \mathcal{P}_{J}^{J_{12},J_{34},J^+_{12}}(w) \times F(u,v) \, .
\end{equation}
The function, $F(u,v)$, depending solely on conformal data is then identified
to the contribution of spin $J$ to the superconformal blocks in equation
\eqref{superconformalBlockHarmonics}, and the harmonics are given by
\begin{equation}\label{hypergeomfun}
	\mathcal{P}_{J}^{J_{12},J_{34},J^+_{12}}(w) = c_J\,w^{-J-(a_1-J_{12}^+)} (1+w)^{-a_2} \, _2F_1\left(-(J+J_{12}),-(J+{J_{34})};-2 J;-w\right)\,. 
\end{equation}
In the main text, we drop the superscript $J_{12}^+$ due to our choice of convention. Notice that the dependence on $a_1, a_2$ drops in the full four-point function due to \eqref{RsymmetryFrame}.

Let us close this appendix by comparing two convenient choices of conventions
that have appeared in the literature:

\begin{itemize} \label{parameterchoice}
	\item $a_1 = J_1 + J_2$, $a_2 = J_{34}$. In this case, the prefactor
		takes a form reminiscent of that of the spacetime prefactor,
		$K_4$, 
		\begin{equation}
			K_4^R = 
			\left(
			Y_{12} \right){}^{\frac{k_{1}+k_2}{2}} 
			\left(Y_{34}\right){}^{\frac{k_{3}+k_4}{2}}
			\left(\frac{Y_{14}}{Y_{24}} \right)^\frac{k_{12}}{2}
			\left( Y_{13} \over Y_{14} \right)^{k_{34} \over 2} \, .
		\end{equation}
		In that case, the superconformal Ward identity takes the simple
		form $(\partial_\chi+\varepsilon\partial_w)F|_{\chi=w}=0$. This
		is the choice taken in this work and in the recent literature
		that have computed the superconformal blocks of four
		momentum-map operators via the Ward identity, e.g.
		\cite{Beem:2014zpa,Chang:2017xmr}. 
	\item $a_1 = 2 E$, $a_2 = 0$ with $2E = J_1+J_2+J_3-J_4$. With this
		choice, the hypergeometric function involved in the harmonic
		\eqref{hypergeomfun} can be recast---up to a constant
		$\tilde{c}_J$---into a very compact form in terms of a Jacobi
		polynomial, $P_{n}^{(\alpha,\beta)}$, 
		\begin{equation}
			F(u,v;w) = P_{J+J_{34}}^{(2J_1 - 2E, 2J_2 - 2E)}\left( 1 + {2 \over w} \right) \times F(u,v)\,.
		\end{equation}
		This is the choice used in \cite{Nirschl:2004pa}, but comes at
		the cost of the Ward identity receiving additional terms with
		respect to the other convention. 
\end{itemize}

\section{Casimirs and Jack Polynomials}\label{app:superconformalcasimir}

Deriving the conformal Casimir equation \eqref{superCasimirEquation} follows in
spirit the same steps used in appendix \ref{app:ConformalFrame} to derive
$SU(2)_R$ harmonics, replacing the R-symmetry Casimir by its superconformal
counterpart, \eqref{superCasimirEquation}. We follow the method of \cite{Bobev:2015jxa,Bobev:2017jhk} throughout
this appendix. 

We look for a representation of the Casimir operator on scalar functions of
spacetime. In the case of the bosonic part, the differential operator can be
inferred in a straightforward manner from the embedding formalism
\cite{Dolan:2003hv}. Acting on a bosonic block, the non-supersymmetric Casimir
equation is then:
\begin{equation}
	\mathcal{D}_\text{bos}\, g^{\Delta_{12},\Delta_{34}}_{\Delta,\ell}(z,\bar{z}) = c^\text{bos}_{\Delta,\ell}\,g^{\Delta_{12},\Delta_{34}}_{\Delta,\ell}(z,\bar{z})\,,
\end{equation}
where the eigenvalue is given in equation \eqref{bosonicCasimirEigenvalue} and the differential operator by 
\begin{align}
	\mathcal{D}_\text{bos} &= \mathcal{D}_z+\mathcal{D}_{\bar{z}} +4\varepsilon \frac{z\bar{z}}{z-\bar{z}}\bigg( (1-z)\partial_z -(1-\bar{z})\partial_{\bar{z}} \bigg)\,,\\
	\mathcal{D}_z  &= 2z^2(1-z)\partial^2_z-(2+\Delta_{34}-\Delta_{12})z^2\partial_z+\frac{\Delta_{12}\Delta_{34}}{2}z\,.
\end{align}

To obtain the differential operator associated to the part of the Casimir involving fermionic generators, one needs to derive how supercharges act on fields at a point $x$. 

Using $\mathcal{O}\phi(x) = (\mathcal{O}\phi)(x)+ i x^\mu [P_\mu,\mathcal{O}]\phi(x)$ and Jacobi identities, one finds that on a single scalar field, 
\begin{align}
	S^{\alpha A}\phi(x) &= (S^{\alpha A}\phi)(x) -ix^\mu\varepsilon^{\alpha\beta}\widetilde{\Gamma}^{AB}_\mu(Q_{\beta B}\phi)(x)\,,\nonumber\\
	Q_{\alpha A}\phi(x) &= (Q_{\alpha A}\phi)(x)\,,\\
	C^2_\text{SUSY}\phi(x) &= 4\phi(x) - (Q_{\alpha A}S^{\alpha A}\phi)(x)\,.\nonumber
\end{align}
The factor 4 in the last equation is related to traces over both $\mathfrak{su}(2)_R$ and spinor indices, giving the total number of supercharges. A relation independent of the number of supercharges can be obtained by replace this factor by $N_S/2$.

We now demand that the scalar is a superconformal primary, which by definition is annihilated by any of the conformal supercharges, $S^{\alpha A}\phi=0$. The Casimir operator acting on two scalar superconformal primaries is then found to be
\begin{equation}\label{SUSYCasimirArbitrary}
	C^2_\text{SUSY}\phi_1(x_1)\phi_2(x_2) = 4(\Delta_1+\Delta_2)\phi_1(x_1)\phi_2(x_2)-i(x_{12})^\mu \varepsilon^{\alpha\beta}\widetilde{\Gamma}_\mu^{AB} (Q_{\alpha A}\phi_1)(x_1)(Q_{\beta B}\phi_2)(x_2)\,.
\end{equation}
We see that the Casimir equation \emph{a priori} looks like it will involve the correlation functions of fermionic fields. However, simplifications occur when considering short superconformal multiplets.

To see this, let us now reestablish the R-symmetry indices and focus on $SU(2)_R$. Using standard $SU(2)$ notation, consider that the scalar field is a superconformal primary of a multiplet of type $\mathcal{D}[J]$, $\phi^{M}$ . This means that it transforms in the spin-$J$ representation, with $-J\leq M \leq J$. By definition, highest and lowest weight states are annihilated by half of the supercharges \cite{Minwalla:1997ka,Buican:2016hpb,Cordova:2016emh},
\begin{equation}
	Q_{1A}\phi^J = 0\,,\qquad
	Q_{2A}\phi^{(-J)}= 0\,,\qquad
	\forall\, A\,.
\end{equation}
This leads us to consider two different classes of correlation functions, which we refer to as type (I) and (II):
\begin{equation}
  \text{(I)}:\,\,\left<\phi_1^{J_1}(x_1)\phi_2^{J_2}(x_2)\phi_3^{M_3}(x_3)\phi_4^{M_4}(x_4)\right>\,,\qquad
  \text{(II)}:\,\,\left<\phi_1^{J_1}(x_1)\phi_2^{M_2}(x_2)\phi_3^{J_3}(x_3)\phi_4^{M_4}(x_4)\right>\,.
\end{equation}
For correlation functions of type (I), the second term of equation \eqref{SUSYCasimirArbitrary} vanishes identically and the Casimir reduces to an application of the dilatation operator. The action of $\mathcal{D}_\text{SUSY}$ on a superconformal block is just multiplication.

In type (II), one can use the fact that in the limit $\left|x_{4}\right|\to\infty$ conformal invariance ensures that no information is lost, and the superconformal Ward identity $\left<Q\mathcal{O}\right>=0$ allows to relate the type (II) Casimir equation to a correlation function where supercharges are only applied on the first field, $\left<Q_{\alpha A}Q_{\beta B}\phi_1(x_1)\cdots\phi\right>$. Using the supersymmetry algebra, the supercharges can be converted into a translation generator. Acting on the whole four-point function decomposed into superconformal blocks, the differential operator will also act on the prefactor, $K_4$, cancelling the term proportional to $\Delta_1+\Delta_2$. 

In summary, we get that the contribution of the supersymmetric part of the Casimir acts on a superconformal block as the following differential operators:
\begin{align}
  \text{(I)}: &~\mathcal{D}_\text{SUSY}  = 4(\Delta_1+\Delta_2)\,,\\
  \text{(II)}: &~\mathcal{D}_\text{SUSY}  = 4z(1-z)\partial_z+4\bar{z}(1-\bar{z})\partial_{\bar{z}} - 2(z+\bar{z}) \Delta_{34}\,.
\end{align}

Finally, it is easy to see that when acting with $\mathcal{C}^2_R$ directly on
the projector, the associated differential operator will act simply as multiplication involving the R-charge of the block element, $J$:
\begin{equation}
	\mathcal{D}_R \mathcal{G}_\chi^J(u,v)= -2\varepsilon J(J+1)\mathcal{G}_\chi^J(u,v)\,.
\end{equation}

\subsection*{Jack Polynomials}

When attempting to solve the constraints coming from the Ward identity of the
Casimir equation, it is useful to decompose the bosonic blocks in terms of Jack
polynomials. We collect here a few properties needed to solve the constraints.
The Jack polynomials can be defined through the Gegenbauer polynomials,
$C^{(\alpha)}_n$, 
\begin{equation}\label{defJack}
	P^{(\varepsilon)}_{\lambda_1,\lambda_2}(z,\bar{z}) = \frac{(\lambda_1-\lambda_2)!}{(2\varepsilon)_{\lambda_1-\lambda_2}}(z\bar{z})^{\frac{1}{2}(\lambda_1+\lambda_2)} C^{(\varepsilon)}_{\lambda_1-\lambda_2}\left( \frac{z+\bar{z}}{2(z\bar{z})^{1/2}} \right) \,,
\end{equation}
normalised such that $P_{\lambda_1,\lambda_2}^{(\varepsilon)}(1,1)=1$. They
satisfy a number of relations, compiled in e.g. \cite{Dolan:2004mu}, following
in part from the properties of the Gegenbauer polynomials. We note that
notation for the coordinates, $z\,,\chi$, in \cite{Dolan:2004mu} and here is
swapped, as we follow conventions that have become common in the recent
literature. 

Under derivation, the Jack polynomial satisfy
\begin{equation}
	z\, \partial_z P^{(\varepsilon)}_{\lambda_1,\lambda_2}(z,\bar{z}) = 
	\frac{\lambda_1+\lambda_2}{2}P^{(\varepsilon)}_{\lambda_1,\lambda_2}(z,\bar{z}) 
	+\frac{z-\bar{z}}{4z\bar{z}}\frac{(\lambda_1-\lambda_2)(\lambda_1-\lambda_2+2\varepsilon)}{1+2\varepsilon}P^{(\varepsilon+1)}_{\lambda_1,\lambda_2+1}(z,\bar{z})
	\,.
\end{equation}
Useful identities involving a Jack polynomial and the two variables are
\begin{align}
	(z+\bar{z}) P^{(\varepsilon)}_{\lambda_1,\lambda_2}(z,\bar{z}) =& \frac{\lambda_1-\lambda_2+2\varepsilon}{\lambda_1-\lambda_2+\varepsilon} P^{(\varepsilon)}_{\lambda_1+1,\lambda_2}(z,\bar{z}) + \frac{\lambda_1-\lambda_2}{\lambda_1-\lambda_2+\varepsilon} P^{(\varepsilon)}_{\lambda_1,\lambda_2+1}(z,\bar{z}) \,,\\
	(z-\bar{z})^2 P^{(\varepsilon+1)}_{\lambda_1,\lambda_2}(z,\bar{z}) =& \frac{2(1+2\varepsilon)}{\lambda_1-\lambda_2+1+\varepsilon}\left( P^{(\varepsilon)}_{\lambda_1+2,\lambda_2}(z,\bar{z}) - P^{(\varepsilon)}_{\lambda_1+1,\lambda_2+1}(z,\bar{z}) \right)\,.
\end{align}
Notice that the second equation involves a Jack polynomial with argument $(\varepsilon+1)$, which allows for the derivative to be rewritten with expression involving only $(\varepsilon)$,
\begin{equation}
	z\partial_z P^{(\varepsilon)}_{\lambda_1,\lambda_2} = 
	\frac{\lambda_1+\lambda_2}{2} P^{(\varepsilon)}_{\lambda_1,\lambda_2} + 
	\frac{(\lambda_1-\lambda_2)(\lambda_1-\lambda_2+2\varepsilon)}{2(z\bar{z})(z-\bar{z})(\lambda_1-\lambda_2+\varepsilon)} \left(P^{(\varepsilon)}_{\lambda_1+2,\lambda_2+1} - P^{(\varepsilon)}_{\lambda_1+1,\lambda_2+2} \right)\,.
\end{equation}
Finally, defining the operator $\mathcal{D}^{(n)} = z^n\partial_z+\bar{z}^n\partial_{\bar{z}}$, one finds 
\begin{align}
	\mathcal{D}^{(1)} P^{(\varepsilon)}_{\lambda_1,\lambda_2}(z,\bar{z}) & = (\lambda_1+\lambda_2) P^{(\varepsilon)}_{\lambda_1,\lambda_2}(z,\bar{z})\,,\\ 
	\mathcal{D}^{(2)} P^{(\varepsilon)}_{\lambda_1,\lambda_2}(z,\bar{z}) & = \frac{\lambda_1(\lambda_1-\lambda_2+2\varepsilon)}{(\lambda_1-\lambda_2+\varepsilon)} P^{(\varepsilon)}_{\lambda_1+1,\lambda_2}(z,\bar{z}) + \frac{(\lambda_2-\varepsilon)(\lambda_1-\lambda_2)}{(\lambda_1-\lambda_2+\varepsilon)}P^{(\varepsilon)}_{\lambda_1,\lambda_2+1}(z,\bar{z})\,.
\end{align}

\section{List of Coefficients for Superconformal Blocks}\label{app:resultCoefficients}

We collect in this appendix the list of coefficients,
$f^J_{\tilde{\Delta},\tilde{\ell}}$, of the superconformal blocks appearing in
the decomposition of four-point functions of four $\mathcal{D}$-type
multiplets. As commented in section \eqref{sec:solvingConstraints}, multiplets
of type $\mathcal{A}$ and $\mathcal{C}$ do not appear in the decomposition.

\subsubsection*{$\mathcal{D}$-Type Multiplets}
The $\mathcal{D}$-type multiplets contain only three states that contribute to
the superconformal blocks. Setting the coefficient of the primary to one, the
other two are given by:
\begin{align}
	f_{2 \epsilon  J_R+1,1}^{J_R-1} &=  \frac{\epsilon (J_R+J_{12})(J_R+J_{34})}{J_R (2 \epsilon J_R+1)}\,,\nonumber\\
f_{2 \epsilon  J_R+2,0}^{J_R-2} &= \frac{\epsilon^2(J_R+J_{12})(J_R+J_{34})(J_R+J_{12}-1)(J_R+J_{34}-1) }{2 (J_R-1) (2 J_R-1) (2 \epsilon  J_R+1) (2 \epsilon  J_R-\epsilon +1)}\,.
\end{align}

\subsubsection*{$\mathcal{B}$-Type Multiplets}
$\mathcal{B}$-type multiplets have fixed conformal dimension, $\Delta =
2\varepsilon J_R+\ell +2\varepsilon$, and the coefficients are found to be:

\begin{align}\label{ResultBType}
	f_{\Delta+1,\ell-1}^{J_R-1} &= \frac{\ell \varepsilon  (J_{12}+J_R) (J_{34}+J_R)}{2 J_R (2 J_R \varepsilon +1) (\ell+\epsilon )}\,, \nonumber\\
	f_{\Delta +1,\ell+1}^{J_R-1} &= \frac{(J_{12}+J_R) (J_{34}+J_R) (l+2 \varepsilon ) (2 J_R \varepsilon +\ell+2 \varepsilon )}{2 J_R (2 J_R+1) (\ell+\varepsilon ) (2 J_R \varepsilon +\ell+\varepsilon +1)}\,,\nonumber\\
	f_{\Delta +1,\ell+1}^{J_R} &= -\frac{J_{12} J_{34} (\ell+2 \varepsilon ) (2 J_R \varepsilon +\ell+2 \varepsilon )}{2 J_R (J_R+1) (J_R \varepsilon +\ell+\varepsilon ) (J_R \varepsilon +\ell+\varepsilon +1)}\,,\nonumber\\
	f_{\Delta +1,\ell+1}^{J_R+1}&= \frac{(-J_{12}+J_R+1) (-J_{34}+J_R+1) (\ell+2 \epsilon )}{2 (J_R+1) (2 J_R+1) (\ell+1)}\,\nonumber,\\
	f_{\Delta +2,\ell}^{J_R-2} &= \frac{\varepsilon  (J_{12}+J_R-1) (J_{12}+J_R) (J_{34}+J_R-1) (J_{34}+J_R) (2 J_R \varepsilon +\ell+2 \varepsilon )}{4 (J_R-1) J_R (2 J_R-1) (2 J_R \varepsilon +1) (2 J_R \varepsilon +\ell+\varepsilon +1)}\,,\nonumber\\
	f_{\Delta +2,\ell}^{J_R-1} &= -\frac{J_{12} J_{34} \varepsilon  (J_{12}+J_R) (J_{34}+J_R) (\ell+2 \varepsilon ) (2 J_R \varepsilon +\ell+1) (2 J_R \varepsilon +\ell+2 \varepsilon )}{4 (J_R-1) J_R^2 (2 J_R \varepsilon +1) (J_R \varepsilon +\ell+\varepsilon ) (J_R \varepsilon +\ell+\varepsilon +1) (2 J_R \varepsilon +\ell+\varepsilon +1)}\,,\nonumber\\
	f_{\Delta +2,\ell}^{J_R}    &= \frac{\varepsilon  (J_R^2-J_{12}^2) (J_R^2-J_{34}^2) (J_{34}+J_R) (\ell+2 \varepsilon ) (2 J_R \varepsilon +\ell+1) (2 J_R \varepsilon +\ell+2 \varepsilon )}{4 J_R^2 (2 J_R-1) (2 J_R \varepsilon +1) (\ell+\varepsilon +1) (2 J_R \varepsilon +\ell+\varepsilon ) (2 J_R \varepsilon +\ell+\varepsilon +1)}\,,\nonumber\\
	f_{\Delta +2,\ell+2}^{J_R} &= \frac{(\ell+2 \varepsilon ) (\ell+2 \varepsilon +1)}{4 (\ell+1) (\ell+\varepsilon +1)}\times\nonumber\\
	&\quad\frac{ (A^- \varepsilon +\ell+1) (A^+ \varepsilon +\ell+1) (B^- \varepsilon +\ell+1) (B^+ \varepsilon +\ell+1) (2 (J_R+1) \varepsilon +\ell)}{ (J_R \varepsilon +\ell+\varepsilon +1)^2 (2 J_R \varepsilon +\ell+\varepsilon +1) (2 (J_R+1) \varepsilon +2 \ell+1) (2 (J_R+1) \varepsilon +2 \ell+3)}\,\nonumber,\\
f_{\Delta +3,\ell+1}^{J_R-1} &=\frac{\varepsilon  (J_{12}+J_R) (J_{34}+J_R) (\ell+2 \varepsilon )(A^- \varepsilon +\ell+1) (A^+ \varepsilon +\ell+1)}{8 J_R (l+1) (2 J_R \varepsilon +1)(J_R \varepsilon +l+\varepsilon +1)^2 (2 J_R \varepsilon +\ell+\varepsilon +1)^2 }\times\nonumber\\
	&\quad\frac{ (B^- \varepsilon +\ell+1) (B^+ \varepsilon +\ell+1) (2 J_R \varepsilon +\ell+2) (2 (J_R+1) \varepsilon +\ell) (2 (J_R+1) \varepsilon +\ell+1)}{ (2 J_R \varepsilon +\ell+\varepsilon +2) (2 (J_R+1) \varepsilon +2 \ell+1) (2 (J_R+1) \varepsilon +2 \ell+3)}\,,
\end{align}
where we defined $A^\pm = J_R+1\pm J_{12},\, B^\pm = J_R+1\pm J_{34}$ and set the coefficient of the primary to one.
\subsubsection*{Long Superconformal Multiplets}

Long multiplets have unconstrained conformal dimensions above the unitarity
bound, and the superconformal blocks receive contributions from 20 different
states, in addition to the superconformal primary, whose coefficient we set to
one by convention.

At level one, the coefficients are given by
\begin{align}\label{longMultipletLevel1}
	f_{\Delta +1,\ell-1}^{J_R-1} =& \frac{\ell (J_{12}+J_R) (J_{34}+J_R) (\Delta-\ell +2\varepsilon J_R )}{2 J_R (2 J_R+1) (\ell+\varepsilon ) (\Delta-\ell+2 +2 (J_R-1) \varepsilon )}\nonumber\\
	f_{\Delta +1,\ell+1}^{J_R-1}=&\frac{(J_{12}+J_R) (J_{34}+J_R) (\ell+2 \varepsilon ) (\Delta +2 (J_R+1) \varepsilon +\ell)}{2 J_R (2 J_R+1) (\ell+\varepsilon ) (\Delta +2 J_R \varepsilon +\ell+2)}\,\nonumber\\
	f_{\Delta +1,\ell-1}^{J_R} =& (-1)\frac{J_{12} J_{34} \ell (\Delta -\ell +2 J_R \varepsilon) (\Delta -\ell -2 (J_R+1) \varepsilon )}{2 J_R (J_R+1) (\ell+\varepsilon ) (\Delta -\ell-2 \varepsilon +2) (\Delta -\ell-2 \varepsilon )}\,,\nonumber\\
	f_{\Delta +1,\ell+1}^{J_R} =&(-1)\frac{J_{12} J_{34} (\ell+2 \varepsilon ) (\Delta + \ell -2 J_R \varepsilon ) (\Delta+\ell +2\varepsilon (J_R+1)  )}{2 J_R (J_R+1) (\Delta +\ell) (\Delta +\ell+2) (\ell+\varepsilon )}\,,\nonumber\\
	f_{\Delta +1,\ell-1}^{J_R+1} =&
	\frac{A^-B^- \ell (\Delta -\ell -2\epsilon (J_R+1))}{2 (J_R+1) (2 J_R+1) (\ell+\epsilon ) (\Delta -\ell+2 -2\epsilon (J_R+2) )}\,,\nonumber\\
	f_{\Delta +1,\ell+1}^{J_R+1} =& \frac{A^- B^- (\ell+2 \epsilon ) (\Delta+\ell -2\epsilon J_R  )}{2 (J_R+1) (2 J_R+1) (\ell+\epsilon ) (\Delta +\ell+2-2\epsilon (J_R+1) )}\,.
\end{align}

At higher level, the expressions become increasingly complex. For instance, the simplest expressions at level two are 
\begin{align}\label{longMultipletLevel2}
	f_{\Delta +2,\ell}^{J_R-2} =& \frac{(J_{12}+J_R-1) (J_{12}+J_R) (J_{34}+J_R-1) (J_{34}+J_R) (\Delta +2 J_R \varepsilon -\ell) (\Delta +2 (J_R+1) \varepsilon +\ell)}{4 (J_R-1) J_R (2 J_R-1) (2 J_R+1) (\Delta +\ell+2 +2\varepsilon J_R ) (\Delta -\ell+2 +2 \varepsilon (J_R-1))}\,,\nonumber\\
	f_{\Delta +2,\ell}^{J_R+2}=&\frac{A^-(A^-+1) B^-(B^-+1) (\Delta -2 J_R \varepsilon +\ell) (\Delta -2 J_R \varepsilon -\ell-2 \varepsilon )}{16 (J_R+1)_2 \left(J_R+\frac{1}{2}\right)_2 (\Delta -2 J_R \varepsilon -\ell-4 \varepsilon +2) (\Delta -2 J_R \varepsilon +\ell-2 \varepsilon +2)}\,.
\end{align}
Note that many of them are related to one another by the relation explained in section \ref{sec:Results}. An exhaustive list of the coefficients for the long multiplets can be found in the \texttt{Mathematica} file attached with the arXiv submission.


\bibliography{references}{}
\bibliographystyle{JHEP} 


\end{document}